\documentclass[sigconf]{acmart}

\AtBeginDocument{%
  }

\copyrightyear{2025}
\acmYear{2025}
\setcopyright{rightsretained}
\acmConference[ASIA CCS '25]{ACM Asia Conference on Computer and Communications 
Security}{August 25--29, 2025}{Hanoi, Vietnam}
\acmBooktitle{ACM Asia Conference on Computer and Communications Security (ASIA CCS '25), August 25--29, 2025, Hanoi, Vietnam}
\acmDOI{10.1145/3708821.3710833}
\acmISBN{979-8-4007-1410-8/25/08}

\usepackage{tikz}
\usepackage{amsmath}
\usepackage{filecontents}
\usepackage{subfig}
\usepackage{algorithmic}
\usepackage{graphicx}
\usepackage{textcomp}
\usepackage{xcolor}
\usepackage{xurl}
\usepackage{balance}
\usepackage{algorithm}
\usepackage{multirow}
\usepackage[english]{babel}
\usepackage{blindtext}
\usepackage{graphicx}
\usepackage{pifont}
\usepackage{subfig}

\newcommand{\ignore}[1]{}

\usepackage{array}
\newcommand{\lov}{{LOV}}
\newcolumntype{P}[1]{>{\centering\arraybackslash}p{#1}}
\usepackage{comment}

\usepackage{filecontents}

\begin{document}


\title{Learning to Identify Conflicts in RPKI}
\ignore{
\author{Haya Schulmann and Shujie Zhao}
    \affiliation{%
      \institution{Fraunhofer Institute for Secure Information Technology SIT}
        \country{}
        }
}

\author{Haya Schulmann}
    \affiliation{%
      \institution{Goethe-Universität Frankfurt, ATHENE}
       \country{}
        }

\author{Shujie Zhao}
    \affiliation{%
      \institution{Fraunhofer SIT, ATHENE}
      \country{}
        }

\begin{abstract}
The long history of misconfigurations and errors in RPKI indicates that they cannot be easily avoided and will most probably persist also in the future. These errors create conflicts between BGP announcements and their covering ROAs, causing the RPKI validation to result in status {\tt invalid}. Networks that enforce RPKI filtering with Route Origin Validation (ROV) would block such conflicting BGP announcements and as a result lose traffic from the corresponding origins. Since the business incentives of networks are tightly coupled with the traffic they relay, filtering legitimate traffic leads to a loss of revenue, reducing the motivation to filter invalid announcements with ROV.

In this work, we introduce a new mechanism, \lov, designed for whitelisting benign conflicts on an Internet scale. The resulting whitelist is made available to RPKI supporting ASes to avoid filtering RPKI-invalid but benign routes. Saving legitimate traffic resolves one main obstacle towards RPKI deployment. We measure live BGP updates using \lov\ during a period of half a year and whitelist 52,846 routes with benign origin errors.


\ignore{
Errors in RPKI deployments and in the announced BGP routes can lead to route leaks and can result in filtering legitimate traffic. Although studies show that both types of errors are common, their number does not decrease indicating that errors are not trivial to avoid.\\
\indent In this work we develop a mechanism, we call \lov, to differentiate between benign errors and real BGP incidents. \lov\ uses machine learning to identify misconfigurations in RPKI and in BGP routes. On the one hand \lov\ identifies benign errors that cause conflicts between RPKI authorizations of ownership and announced BGP routes and prevents legitimate traffic from being blocked. On the other hand \lov\ blocks harmful hijacks and route leaks. \lov\ offers the benefits of RPKI validation, while resolving one of the main obstacles towards deployment of RPKI: {\em benign errors}, hence paving the way for deployment of path validation mechanisms.\\ 
\indent We integrate \lov\ into a Routinator implementation of RPKI validator and evaluate \lov\ on a real-world dataset with major BGP hijacks and route leak incidents, demonstrating detection accuracy of 99.9\%. We deploy \lov\ on the Internet, and report evaluation results over a period of half a year: \lov\ found that 81.1\% of the conflicting routes were caused by benign misconfigurations, which would have been dropped by RPKI validators disconnecting legitimate origins.
Alternately, \lov\ detected 344 hijacks and 451 route leak incidents, of which 36 were confirmed by corresponding network operators in a survey that we conducted.
We provide our code and datasets at {\small \path{https://github.com/zsjstart/LOV}}.
}
\ignore{
In addition to identifying and “saving” benign conflicts,  \lov\ also extends ROV in identifying and filtering route leaks. Route leaks are not filtered by RPKI since route leaks also do not introduce more specific routes and neither do they alter the origin of the route, hence the origin in route leaks is the legitimate owner of the announced prefix. Although route leaks introduce a problem to Internet stability and occur often, there is currently no efficient solution to this. We show experimentally how \lov\ detects and blocks route leaks.
}

\ignore{
RPKI has been proposed for securing BGP for more than ten years; still, only a limited number of networks adopt ROV.
Without widespread adoption of ROV, RPKI does not provide any security benefits.
One of the main factors hindering ROV's wide adoption is invalid ROAs.
ROA misconfigurations result in conflicting ROA and BGP announcement pairs, which appear similar to prefix hijacks. ROV will filter such conflicts causing routers to block legitimate traffic.
Besides, ROV cannot detect path anomalies because the AS number at the end of the path is indeed the legitimate owner of the prefix announced.
These shortcomings make network operators less motivated to deploy RPKI in practice.

Our work aims to promote the practical deployment of RPKI.
In this work, we present a new extension of ROV, called LOV, that builds the validation process in an intelligent mode using machine learning technologies.
LOV extends ROV with the following properties:
(1) LOV can rescue conflicts due to erroneous ROAs. (2) LOV can identify path anomalies.
(3) LOV can whitelist benign routing anomalies.

We evaluated LOV on the real-world dataset, showing that LOV has a low false positive rate of 0.6\%, and a high true positive rate of 98.6\%, outperforming ROV. Also, we show that LOV can save BGP announcements with ROA misconfigurations from being dropped.
With LOV, we studied 12 major BGP incidents, showing that LOV can effectively combat BGP hijacks and route leaks.
Besides, we measured historical BGP updates using LOV to characterize conflicts and malicious events.
Finally, we evaluated the impact of LOV on Internet reachability and BGP convergence through simulations, showing that LOV is readily deployable.
}
\end{abstract}



\maketitle

\begin{sloppypar}

\section{Introduction} \label{sec:lov_overview}
Border Gateway Protocol (BGP) \cite{rfc4271} serves as the inter-domain routing system of the Internet, enabling routers to exchange information about the reachability of prefixes within their Autonomous Systems (ASes). However, ASes can originate prefixes that they do not legitimately own, leading to routing misconfigurations or intentional hijacks. Such incidents allow attackers to redirect traffic through their networks, facilitating eavesdropping, blackholing of hijacked traffic, or dissemination of spam \cite{ballani2007study,vervier2015mind,vervier2013spamtracer}.

\textbf{RPKI to secure Inter-domain routing.}
To combat these issues, the Internet Engineering Task Force (IETF) has standardized the Resource Public Key Infrastructure (RPKI) \cite{RPKI}, which binds prefixes to AS numbers (ASNs) and their corresponding cryptographic public keys via Route Origin Authorizations (ROAs).
Route Origin Validation (ROV) \cite{huston2012validation} validates BGP announcements by ensuring that the origin AS matches the authoritative AS specified in the ROAs and that the prefix length is within the allowed range.
By enforcing ROV, ASes can filter out invalid BGP announcements, thereby preventing the propagation of unauthorized routes.

Qrator Labs recently reported that the number of BGP hijack incidents declined from 9,022 in Q2 2022 to 7,595 in Q2 2023 \cite{Qrator}. They credited this drop to the growing adoption of RPKI among ASes.
Despite experiencing some growth, the deployment of RPKI is advancing at a slow pace.
As of January 2024, only 33\% of 413 major network operators have implemented ROV to prevent illegitimate BGP routes \cite{cloudflare}. Li et al. \cite{li2023rovista} discovered in 2023 that merely 12.3\% of ASes have fully employed ROV.

\textbf{Obstacles to RPKI adoption.}
The limited adoption of ROV poses a potential risk of collateral damage to networks currently deploying this technology, preventing them from obtaining sufficient security benefits \cite{gilad2016we,hlavacek2020disco,morillo2021rov++,ixp:HlavacekSVW23}. 
One of the reasons behind the hesitation to enforce ROV is the concern about the loss of legitimate traffic due to misconfigurations in ROAs \cite{hlavacek2020disco,hlavacek2022smart}. Manually issuing ROAs often leads to errors in the ROAs \cite{hlavacek2020disco}. For instance, network operators may incorrectly configure the maximum length of prefixes. These erroneous ROAs conflict with legitimate BGP announcements and result in an invalid ROV validation status, and consequently, the corresponding legitimate BGP announcements are blocked by ROV-filtering networks.
Such an origin conflict is known as ``benign conflict'' since it is caused by misconfigurations and not by actual hijacks. Filtering benign conflicts due to erroneous ROAs adversely affects the reachability of the prefixes covered by those ROAs.
According to the NIST RPKI monitor \cite{nist2024}, the quantity of erroneous ROAs grows proportionally as the number of prefixes covered by ROAs increases.
As the enforcement of ROV filtering continues to rise among networks, the impact of benign conflicts on network reachability will exacerbate.
Since BGP operations are driven by economic incentives, the loss of legitimate traffic can directly result in a loss of revenue.

\textbf{Improving ROV implementation to promote RPKI deployment.}
Previous studies, such as \cite{hlavacek2020disco,hlavacek2022smart}, primarily focus on errors in ROAs.
DISCO \cite{hlavacek2020disco} explores methods for mitigating human errors in ROA issuance, while SROV \cite{hlavacek2022smart} delves into identifying benign conflicts resulting from erroneous ROAs. However, our study reveals that intricate routing policies or their changes—such as multiple origin ASes (MOAS), routing shifts, and traffic engineering—can also lead to benign conflicts. These conflicts are often not due to ROA errors, which present additional challenges for resolution.

In our measurements, benign conflicts are observed to be both persistent and widespread compared to hijacks, with thousands of routes potentially affected daily, persisting over weeks or even months.
Our study reveals that approximately 79\% of the ROV-invalid routes per day are attributed to benign conflicts with ROAs, with 26K unique benign conflicts recurring within an average span of less than 14 days. 
Sacrificing a significant volume of legitimate traffic to prevent a relatively smaller number of hijacks would be both impractical and detrimental to overall network operations.

SROV, as an extension of ROV, using a heuristic approach based on route duration to analyze BGP announcements, may introduce BGP convergence delays.
Additionally, this approach has a high rate of false negatives (i.e., misidentifying hijacks as benign routes), potentially leading to failures in preventing actual BGP hijacks. In practical terms, striking a balance in ROV implementation—ensuring security and maintaining network performance—is essential.
As a result, new mechanisms need to be explored to resolve these issues.
Our work addresses this balance by preventing the discard of BGP announcements with benign conflicts while still upholding the intended function of ROV. By achieving this, our work has the potential to support the broader adoption of RPKI and its ROV.

\textbf{Learning Origin Validation (\lov).}
This work introduces Learning Origin Validation (\lov), designed for whitelisting benign conflicts on an Internet-wide scale. The generated whitelist is offered to the ASes that employ ROV to validate RPKI-invalid routes. \lov\ matches the RPKI-invalid routes against the whitelist. If a match is found, the route is considered benign, and border routers refrain from blocking it.

Notably, we utilize a whitelist of benign conflicts rather than a blacklist of hijacks, as ROV is already capable of perfectly preventing hijacks involving origins that conflict with ROAs.
Additionally, the transient and rapidly changing nature of anomalous routes makes maintaining an up-to-date blacklist challenging.
The whitelist targets benign conflicts, which are typically long-lived, thereby minimizing the synchronization issues with routers, that often arise when using a blacklist to address short-lived BGP hijacks.
Consequently, using a whitelist provides a more effective and reliable approach.

In contrast to SROV, \lov\ aims to provide a \textit{high-quality} whitelist for networks enforcing ROV. To achieve this, \lov\ incorporates three key mechanisms: a \textit{ML-classifier} that leverages machine learning to identify benign conflicts in the initial stage; a \textit{Post-analyzer}, which monitors changes in the global visibility of origin ASes to verify potential hijacks detected by the ML-classifier, addressing misidentifications; and a \textit{Quarantine} mechanism, which ensures the reliability and trustworthiness of the whitelist through multiple review strategies. An overview of \lov\ is offered in Section \ref{sec:lov_overview}.

\lov\ is designed to identify various types of benign conflicts without assuming human error or attempting to eliminate them. It operates on a public web server, regularly providing a whitelist to ASes that enforce ROV, without requiring significant changes to existing ROV deployments. The process of downloading, installing, and maintaining this whitelist is straightforward and incurs minimal computational overhead compared to potential ROV extensions like SROV. As a result, this approach helps avoid a negative impact on BGP convergence.

{\bf Contributions.} We make the following contributions:

We introduce \lov, a new mechanism designed to offer whitelisted benign conflicts to ROV-enforcing ASes, preventing legitimate announcements involving benign conflicts from loss (Section \ref{sec:lov_overview}).

We present features utilized to categorize RPKI-invalid routes into benign conflicts and BGP hijacks (Section \ref{sec:lov_features}).

We employ five ML models to implement the binary classifier, demonstrating that the Random Forest model achieves optimal performance with 95\% accuracy in identifying benign conflicts and $\sim$100\% accuracy in detecting BGP hijacks (Section \ref{subsec:groundtruth}).

We conduct an extensive measurement of live BGP data using \lov\ over a six-month period, ultimately whitelisting 52,846 benign conflicts on the Internet, and provide insights into the underlying causes of benign conflicts, presenting four additional factors, aside from human errors (Section \ref{sec:measure}).

We address key questions regarding the implementation and deployment of \lov\ (Section \ref{sec:questions}), and discuss future research questions, potential improvements, and prospective uses (Section \ref{sec:lov_future}).


\ignore{
 \paragraph{Organization} The rest of the paper is organized as follows. Section \ref{sec:background} introduces background and preliminaries. Related work is presented in Section \ref{sec:relatedwork}.
 Section \ref{sec:features} explains the features used for distinguishing between benign conflicts and BGP hijacks. Section \ref{sec:data_sources} describes data sources we use for feature computation. In Section \ref{sec:implementation} we implement \lov\
 and validate its performance. A long-period live measurement is presented in Section \ref{sec:measure}. We compare with BGPmon in Section \ref{sec:compare} and discuss \lov\ in Section \ref{sec:discussion}. Finally, we conclude in Section \ref{sec:conclusion}.
 }

\ignore{
Our individual technical contributions can be summarized as follows:
 $\bullet$ {\em Mechanism for filtering hijacks and saving benign conflicts.} We extend the Routinator implementation\footnote{\url{https://github.com/NLnetLabs/routinator}} with modules we developed for differentiating benign conflicts from hijacks. Our mechanism, which we call Learning Origin Validation (\lov), is based on a machine learning (ML) classifier. Compared to heuristic approaches, which often rely on manually defined thresholds that may not be adaptable to changing conditions, ML algorithms can automatically identify patterns in the input features, resulting in better classification performance; in addition, the ML-based approach allows for automatic updating of the model with new data, making it a more scalable and adaptive solution for detecting routing anomalies in today's dynamic and complex Internet environments. We demonstrate this with extensive experimental evaluation on heuristically derived datasets on a simulation platform as well as with a longitudinal evaluation on the Internet. 

 $\bullet$ {\em Evaluation on ground truth datasets.} We perform evaluation of \lov\ on BGP routes we collect on the Internet. We collect BGP announcements with benign conflicts, which were visible between May 2022 and June 2022 from Routeviews \cite{routeviews}, resulting in a dataset of 9,223 long-lived and harmless benign conflicts. We also collecte BGP incidents between December 2021 and September 2022 from BGPmon (BGP monitoring service) \cite{BGPMon}. This dataset consists of 415 BGP hijacks. In addition, we collect 8 BGP incidents with significant global impact between 2020 and 2022 from another well-known data source Qrator \cite{Qrator}, resulting in 10,068 hijacks.

For training the DT-classifier, we use a dataset with 2,000 benign conflicts and 415 anomalous instances identified by BGPmon. We then constructed a holdout dataset by combining the remaining 7,223 benign conflicts with hijacks collected from Qrator. This holdout set was used to evaluate the performance of \lov's DT-classifier and post-analyzer. Our experimental results demonstrate that \lov\ saved 95\% of the benign conflicts in the holdout set, while achieving close to 100\% accuracy in identifying hijacks. These results demonstrate that \lov\ can effectively identify legitimate traffic in cases of erroneous ROA without sacrificing the effectiveness of ROV in blocking BGP hijacks.

$\bullet$ {\em Impact of \lov\ on connectivity and BGP convergence.} We evaluate the impact of \lov\ on a simulated platform on Internet connectivity and BGP convergence on 560 networks. Our simulations involve 9,223 routes with benign conflicts and 1000 routes with hijacked prefixes. The results show that compared to the conventional ROV approach, deploying \lov\ significantly reduces the loss of legitimate routes, thereby improving Internet connectivity. Additionally, the likelihood of undetected hijack attempts in the presence of \lov\ is low, maintaining a comparable level of Internet reachability with ROV. Furthermore, the computational overhead introduced by \lov\ is minimal and does not significantly impact the convergence time of BGP advertisements (i.e., BGP convergence).

 $\bullet$ {\em \lov\ shows most conflicts are caused by errors.} Between October 1st, 2022, and March 31st, 2023, we conduct a long-term measurement of live BGP data with \lov\ using BGPStream. Our analysis revealed that, on a daily average, 79\% of the routes that violated ROAs were due to benign misconfigurations. These benign conflicts affected 61\% of the measured prefixes and were traced back to 17\% of the measured ASes. Our findings indicate that benign conflicts are both persistent and prevalent on the Internet. Further analysis of the underlying causes of these benign conflicts shows that the predominant factors for errors are: prefix deaggregation (45,941 instances), downward dependencies (12,382 instances), multi-origin ASes (4,524 instances), and dangling ROAs (3,879 instances).

  $\bullet$ {\em \lov\ shows most hijacks are not by origin in BGP paths.} Our analysis between October 1st, 2022, and March 31st, 2023 also identified a total of 119,815 hijacks during the measurement period. Notably, we found that 90\% (108,104) of the hijacks were exported through AS212483, one of the peers feeding BGP data to the route collector used, between October 26th and November 30th, 2022. Our analysis suggests that the majority of the hijacks detected were caused by AS212483 rather than the origins in BGP paths. To the best of our knowledge, this is the first report to document these incidents. 

$\bullet$ {\em Survey of network operators.} We conduct a survey of 194 network operators from networks that were involved in hijacks, and received 11 responses, all of which were aligned with the classification of \lov. Our findings also highlight that BGP hijacks are typically the result of misconfigurations, technical errors, or unintended behavior of network operators, such as routing policy changes or DDoS mitigation. 
}

\ignore{
One of the most effective ways to intercept traffic between a source and a destination is via Border Gateway Protocol (BGP) prefix hijacks \cite{china:telecom,fb:out,turkey:hijack}. Prefix hijacks occur when routers send BGP announcements claiming prefixes that belong to other Internet destinations. Such hijacks often happy due to attacks as well as misconfigurations. Autonomous Systems (ASes) that accept the bogus BGP announcements send traffic for IP addresses inside the hijacked prefix to the attacker instead of the legitimate destination. 

Resource Public Key Infrastructure (RPKI) \cite{RPKI} was designed to authenticate network resources that ASes own with Route Origin Authorizations (ROAs), and to enable routers to validate if BGP announcements that they receive are correct via Route Origin Validation (ROV) filtering. Although RPKI was standardized more than ten years ago, still only about 30\% of the ASes filter bogus BGP announcements with ROV \cite{cloudflare}. However, without widespread adoption of ROV RPKI does not provide any security benefits. In particular, previous work showed that the security benefit against hijacks during partial adoption are negligible \cite{gilad2016we}. One of the main factors that hinders wide adoption of ROV are invalid ROAs. Out of the 40\% of the prefixes currently covered with ROAs, 10\% are invalid due to misconfigurations. 
Most commonly such invalid ROAs result due to incorrect signing, incorrect AS number in the ROA, BGP announcements with more specific IP prefixes than the allowed prefix in the max-len in ROA or changes in the IP address allocation. For instance, consider a provider AS3215 with prefix {\tt 193.2.0.0/15} that delegates some of its IP prefix to its two customers: AS1272 {\tt 193.2.35.0/24} and AS8361 {\tt 193.2.155.0/24}. If the provider AS3215 issues an ROA for its prefix {\tt 193.2.0.0/15} with a max-len of 24 for origin AS 3215, but the customers do not issue ROAs for their prefixes, then the ROV filtering of all the announcements of the customers will result in status invalid and will blocked.

Such misconfigurations result in conflicting ROA and BGP announcement pairs, which appear similar to prefix hijacks. ROV filtering such conflicts causes routers to block legitimate traffic. Misconfigurations in ROAs is one of the main reasons for low adoption of ROV filtering. Despite significance of creating valid ROAs and enforcing ROV and awareness to prefix hijacks, there are still a large number of conflicting ROA an BGP announcement pairs due to misconfigurations. Although invalid prefixes are a long known issue, the number of invalid ROAs increases proportionally to the number of ROAs. We analyze the causes for misconfigurations and find that above 85\% of 
 them are errors, which are difficult to avoid and therefore misconfigurations will most likely always persist. 

In this work we analyze the characteristics of ROAs that conflict with BGP announcements due to misconfigurations in ROAs and evaluate their impact on reachability of the affected networks. We quantify the fraction of the Internet impacted by the invalid ROAs. 

{\bf Contributions.}

We analyze historical BGP data.

We create datasets.

We develop machine learning based classifier to identify conflicts that result due to invalid ROAs.
}

\vspace{-10pt}
\section{RPKI Overview} \label{sec:background}

Routers use BGP packets to exchange reachability information with their neighbors. A BGP announcement includes an IP prefix and the AS path to that prefix, with the source AS always located at the end of the path, typically denoted as a tuple of (AS, prefix, path). 
After processing a BGP announcement, the router prepends its AS number to the AS path and forwards it to the next hop.

{\bf Prefix/BGP hijacks.} Prefix/BGP hijacks occur when an AS announces routes for a prefix it does not legitimately own. The attacker can launch a subprefix hijack by announcing a more specific prefix of the victim's, or a prefix hijack by announcing the same prefix as the victim's.
When routers accept the bogus announcements, they redirect traffic through the hijacking network, enabling the attacker to eavesdrop on communication, blackhole hijacked traffic, or engage in other hostile activities like spamming \cite{ballani2007study,vervier2015mind,vervier2013spamtracer}.
A single prefix hijack refers to the announcement of a specific prefix by an individual hijacking AS, while a hijacking event typically involves a series of hijacks launched by the perpetrator AS.

{\bf RPKI.} The goal of RPKI is to provide authenticated information on prefix ownership on the Internet, which routers can use for making routing decisions in BGP. 
A publication point keeps a finite set of signed authorizations, called ROAs \cite{rfc6480}. 
Each ROA binds IP address blocks to owner ASes via cryptographic signatures and is signed by the secret signing key of the owner of the publication point. 
RPKI facilitates many mechanisms, including the filtering of bogus announcements with Route Origin Validation (ROV) or path validation mechanisms, such as \cite{aspa,rsc,rfc8205}.

{\bf Route Origin Validation (ROV).}\label{sc:rov:validation}
RPKI-based route origin validation, that is ROV, defined in RFC6483 \cite{huston2012validation}, allows BGP routers to verify whether an incoming BGP announcement conforms to ROAs. 
To perform ROV, an AS requires a relying party software installed on a local cache machine to periodically fetch resource certificates and ROA objects from RPKI repositories and validate ROAs. The validated ROAs are cached, and are retrieved by the routers over an RTR protocol for making routing decisions in BGP. 
Given a BGP announcement, the validation proceeds by checking if (1) the IP prefix in the ROA covers the prefix in the BGP announcement, (2) the AS number in the ROA is the same as the origin AS in the BGP announcement, and (3) the prefix length in the BGP announcement is smaller or equivalent to the MaxLength parameter in the ROA. A border router that receives a BGP announcement ($p$, $l$, A) for prefix $p$ with prefix length $l$ from AS A, checks if it has a ROA ($p*$, $l*$, A*) for prefix $p*$ from AS A*, such that: $p*$ covers $p$, A* and A are equivalent, and max length $l* \geq l$. If so, the result of ROV for ($p$, $l$, A) is valid. Otherwise, invalid.
If no matching ROA can be found, the status of the BGP announcement is then set as unknown.

\vspace{-10pt}
\section{Related Work} \label{sec:relatedwork}
A number of approaches were proposed for detecting BGP hijacks.
Karlin et al. \cite{karlin2006pretty} introduced Pretty Good BGP (PGBGP), which uses a window of historical BGP updates to quarantine suspicious announcements and delay their adoption. However, this approach can incur significant resource overhead due to the need to manage large volumes of historical data. Additionally, the quarantine mechanism can delay the acceptance of legitimate announcements with benign conflicts.
Schlamp et al. \cite{schlamp2016heap} inferred legitimate business relationships between the hijacking and victim ASes using IRR databases, but encountered potential challenges due to the possibility of outdated or conflicting information within the databases. Tools such as iSPY \cite{zhang2008ispy} and Argus \cite{shi2012detecting} detect prefix hijacks by relying on the concept that polluted ASes fail to reach the victim's network. However, they require real-time probing by the deploying network, which may impose non-negligible computational or communication overhead.
Other studies, such as \cite{deshpande2009online, karimi2019border, testart2019profiling, hlavacek2022smart}, rely on features like AS path length, the number of BGP updates, the longevity of prefix advertisements, and hijack duration. However, AS path length can vary during BGP propagation, and features related to the number, longevity, or duration of BGP announcements may affect the identification of hijacks in their early stages. Consequently, these features may not be suitable for prompt detection.

As a de facto measure to prevent BGP hijacking, the deployment of RPKI started in 2009, and in 2010, Randy Bush had a series of presentations to promote the new standard candidates \cite{bush:rpki:2009}. At that point, there was just a test-hosted RPKI system by RIPE, with no practical implementation of delegated certificate authorities and publication points. In 2011, the production-ready hosted RPKI by RIPE was established, and Deutsche Telekom created ROAs for their prefixes, which de-facto jump-started the RPKI in RIPE region \cite{band:rpki:2011}. Many networks quickly created ROAs for their prefixes, and as of February 2023, there were 140K ROAs, covering approximately 39.3\% of IPv4 prefixes announced in the default-free zone (DFZ)\footnote{The DFZ has a full global BGP routing table.}, with about 8\% being invalid due to errors in ROAs. In contrast to ROAs, the enforcement of ROV progresses much slower. The most recent measurement \cite{hlavacek2022behind} shows that only 27\% of the ASes enforce ROV filtering. The main factor hindering the adoption of ROV is the fear of losing traffic due to erroneous ROAs \cite{gilad2017we,iamartino2015measuring, wahlisch2012towards}. Also, a follow-up study \cite{chung2019rpki} shows that the number of invalid ROAs does not decrease with time but remains between 8\% and 10\%.

Previous studies identified vulnerabilities, errors and misconfigurations in RPKI \cite{stalloris:HlavacekJMSW22,mirdita2022poster,beyond:HlavacekJMSW23,cure:MirditaSVW24,rpki:sok:mirdita25}. A recent proposal showed how to utilize Byzantine consensus to enhance the resilience of RPKI against vulnerabilities and errors \cite{byzrp:FriessMSW24}.

\ignore{
Efforts to encourage broader adoption of RPKI have focused on addressing ROA misconfiguration issues. DISCO \cite{hlavacek2020disco} automates ROA publishing.
It can only prevent human errors in ROA issuance but not handle benign conflicts resulting from complex routing policies such as traffic engineering or MOAS.
SROV \cite{hlavacek2022smart} uses a heuristic approach based on route duration to analyze errors in ROAs. However, this approach may cause BGP convergence delays, such as monitoring new BGP announcements for several weeks, and could potentially weaken ROV's ability to prevent hijacks due to high false negative rates.
}


\ignore{
{\bf Route leaks.} Existing approaches to block route leaks rely on marking routes to avoid leakage or use heuristics and shared routing information to identify valley-free violations. 
RFC9234 \cite{rfc9234} introduces a new BGP path attribute, so-called Only to Customer (OTC), that transit ASes can use for detecting route leaks based on BGP Roles \cite{rfc5492} such as customer, provider, or peer. The OTC attribute is set to the AS number of the advertising AS only when a route is exported to a customer or a peer.
When a provider receives a route with an OTC value set, the route will be discarded to prevent route leaks.
The problem is that deploying OTC requires changing the BGP protocol and the BGP standard and integrating a new configuration parameter called BGP role. This parameter needs to be negotiated using BGP OPEN messages between BGP routers. 
Recently, \cite{aximov2020verification} introduced a new digitally signed RPKI object called Autonomous System Provider Authorization (ASPA), binding a customer AS number and the authorized provider AS number for its announcements. The AS path verification with RPKI checks if the routes received from customers or peers comply with the ASPAs.
Apart from leak prevention, this method can protect against path manipulation attacks. For its deployment ASPA requires fully deployed RPKI and ROV. Therefore, resolving the obstacles towards enforcement of ROV is not only important for blocking prefix hijacks but also for adoption of other path validation mechanisms. In addition, the same challenges that apply to RPKI also apply to ASPA. First, motivating operators to deploy ASPA is challenging, indeed we have not found ASPA deployments in the wild. Second, errors in ASPA objects configurations as well as in validation are inevitable. Therefore, we expect that similar concerns will need to be addressed with deployment of ASPA. 


Other proposals, like \cite{bagnulo2022practicable,MANRS}, use operator registration information in IRR databases to detect route leaks, since the IRR databases are often misconfigured, both approaches have high false positives/negatives rates and are not used in practice.
Systems like Peerlock and Peerlock-lite \cite{mcdaniel2020flexsealing} can be deployed between two BGP neighbors to prevent illegitimately transiting the protected AS by filtering routes that contain an AS number of a large network (e.g., Tier-1 or unauthorized upstreams) on the path. The study in \cite{mcdaniel2020flexsealing} showed that Peerlock/Peerlock-lite was offers protection to large transit networks; thus, the number of protected networks remained low. However, the lack of automated configuration protocol makes Peerlock vulnerable to misconfigurations \cite{mcdaniel2020flexsealing}. Our approach for blocking route leaks does not suffer from the misconfigurations inherent in previous works and it does not require complex configuration. Since \lov\ is an extension of an existing Routinator implementation, networks benefit by simply enforcing ROV. 
}

\ignore{
Some efforts have been delved to improve today's RPKI system.
Publishing ROAs has been a manual process since RPKI was standardized, inevitably leading to unintended human errors \cite{gilad2016we,iamartino2015measuring, wahlisch2012towards}.
Disco \cite{hlavacek2020disco} enables automatic certification of IP address blocks to reduce ROA errors for better RPKI benefit.
As mentioned earlier, only subtle security benefits will be produced without the widespread deployment of RPKI. 
ROV++ \cite{morillo2021rov++} is another extension of ROV that improves security benefits even with low adoption.
Although the authors conducted a detailed evaluation of the simulation platform to show the validity of ROV++, there is no evidence to prove its practicality in the real world.
Differently, our work extends the functionality to ROVs by rescuing conflicts due to wrong ROAs, enabling filtering of advertisements with anomalous paths (e.g., route leaks), and providing post-verification and long-term monitoring to verify routing anomalies and whitelist long-lived and harmless unverified routes. 
}
\ignore{
A number of approaches have been proposed for detection of BGP anomalies or classification of BGP behavior.
Several techniques \cite{zhang2004detection, deshpande2009online} leveraged signatures or statistics of routing updates to detect BGP anomalies.
Differently, to detect unintentional routing anomalous behavior, Lutu et al. \cite{lutu2015bgp} proposed a tool, BGP visibility scanner, which can provide network operators with the status of prefix visibility.
In addition, Teoh et al. \cite{teoh2006bgp} presents BGP Eye, a visualization tool, aiming at real-time detection and root-causes analysis of BGP anomalies.
}

\vspace{-5pt}
\section{LOV Overview} \label{sec:lov_overview}
\begin{figure}[t!]
\centerline{\includegraphics[width=0.8\linewidth]{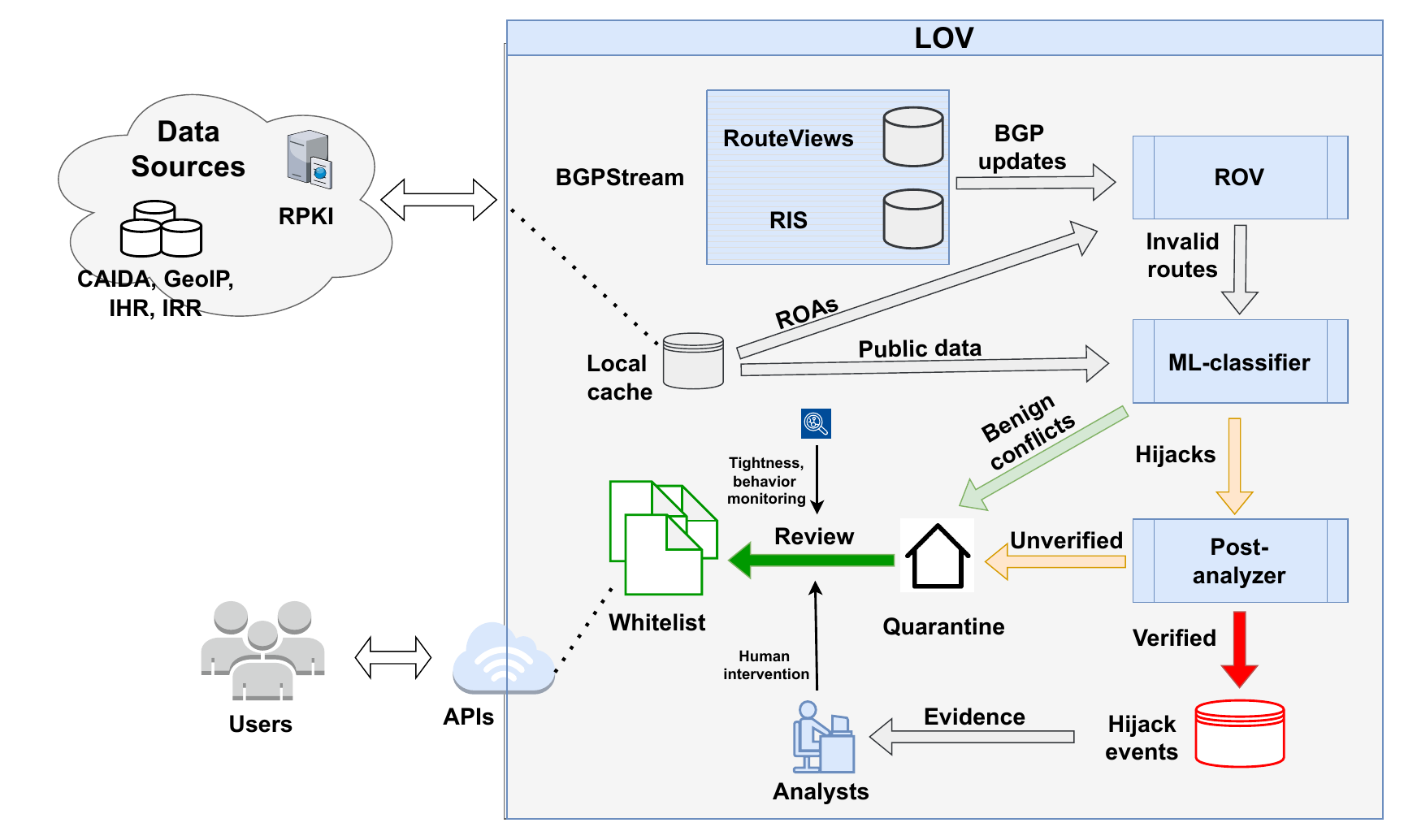}}
\vspace{-10pt}
\caption{\small{The overview of \lov.}}
\label{fig:lov_overview}
\end{figure}

Figure \ref{fig:lov_overview} illustrates an overview of \lov. The core goal of \lov\ is to provide network operators enforcing ROV with a high-quality whitelist, preventing benign conflicts from being filtered out while maintaining ROV's effectiveness in preventing hijacks.
\lov\ operates in the following four steps, outlined below:

\noindent\textit{1) RPKI's ROV: Validates BGP announcements.}
\lov\ utilizes BGPStream \cite{orsini2016bgpstream} to collect live BGP updates from RouteViews and RIS route collectors\footnote{As of writing, there are 39 RouteViews and 26 RIS route collectors.}, and then validate them using RPKI's ROV.

\noindent\textit{2) ML-classifier: Classifies RPKI-invalid routes.}
Legitimate BGP announcements with benign origin conflicts are mistakenly categorized as invalid routes by RPKI's ROV.
\lov\ resolves this issue by employing an ML-based classifier (denoted as ML-classifier) to re-validate the invalid routes.
The ML classifier categorizes an RPKI-invalid route as either a benign conflict or a BGP hijack.
The classification model is trained using a list of features that illustrate the connections between two conflicting origins present in BGP announcements and ROAs (refer to Section \ref{sec:lov_features} for more details). The feature computation relies on public data, including AS relationships, AS organizations, AS geolocations, and AS hegemony values.
\lov\ downloads the necessary data from reliable sources, such as CAIDA \cite{CAIDA}, IHR \cite{IHR}, and IRR \cite{IRR}, and stores it in a local database, ensuring readiness for the ML-based classification. Daily updates are performed to keep the data current.
The features used should enable the ML classifier to effectively identify benign conflicts without affecting ROV's performance in detecting actual hijacks.
In other words, the ML classifier prioritizes the accurate identification of hijacks, even if it means that some benign conflicts may be misclassified as hijacks.

\noindent\textit{3) Post-analyzer: Verifies potential hijacks.}
Since the ML classifier is designed to favor the detection of hijacks, it may sometimes mistakenly classify certain benign conflicts included in invalid routes as hijacks. To address these potential misclassifications, \lov\ employs a post-analyzer to further verify the suspected hijacks.
The post-analyzer examines changes in global AS visibility to identify patterns indicative of BGP hijacking events.
This approach can verify whether the suspected routes indeed originate from BGP incidents or not (refer to Section \ref{subsec:lov_post_analyzer} for more details).
As illustrated in Figure \ref{fig:lov_overview}, the hijacks detected by the ML-classifier are forwarded to the post-analyzer.
Verified routes are recorded as potential BGP hijacking incidents for further analysis, contributing to the enhancement of BGP security. Routes that cannot be confirmed as hijacked might be innocent and are quarantined for further review. 

\noindent\textit{4) Quarantine mechanism: generates a trusted whitelist.}
To ensure the quality of the whitelist, benign conflicts identified by the ML-classifier and unverified routes flagged by the post-analyzer are not immediately added to the whitelist but are instead isolated for further scrutiny, as depicted by ``Quarantine'' in Figure \ref{fig:lov_overview}.
Routes in quarantine are reviewed using various whitelisting strategies, including examining the tightness level between two conflicting origins, continuous behavior monitoring, and human intervention.
The analysis of hijacking events is primarily conducted through email surveys, which provide evidence (e.g., email confirmations) to security analysts for potential human intervention (refer to Section \ref{subsec:quarantine} for more details).

To streamline the whitelist, we extract the originating AS and the announced prefix from the route, creating a unique entry (i.e., a pair of AS and prefix) in the whitelist.
\lov\ updates the whitelist daily. New entries are automatically added to the whitelist, and concurrently, old entries that have not been seen for longer than a month, or that have become RPKI-valid or unknown are purged.
Furthermore, \lov\ provides APIs that allow users to access the whitelist.
\vspace{-10pt}
\section{ML-Classifier} \label{sec:lov_classifier}
This section implements the ML-classifier that is utilized to categorize RPKI-invalid routes into benign conflicts and BGP hijacks. We apply five widely employed supervised learning classification algorithms: Decision Tree (DT), Support Vector Machine (SVM), K Nearest Neighbor (KNN), Random Forest (RF), and Naive Bayes (NB), for validating our approach. The performance of each model is evaluated, and the optimal model is selected as the final classification model.
\vspace{-10pt}
\subsection{Features} \label{sec:lov_features}
We first introduce a set of features that are utilized to train an ML model to distinguish between benign conflicts and BGP hijacks.
Unlike RPKI's ROV, which compares two origins—the origin AS in the prefix announcement and the authoritative AS specified in the ROA—the core idea of our approach to route classification is to identify connections between the two origins.

Intuitively, a BGP announcement is considered benign when the two origins are closely linked.
The features selected cover various associations between two origins, as detailed in Table \ref{tab:features}, along with the data sources used to compute specific features.
Certain features we select are inspired by the previous work \cite{chung2019rpki}, which examines RPKI-invalid announcements. By learning from these features, the ML model captures the degree of tightness between two origins for a given new announcement. The tighter the connection between the two origins, the higher the likelihood of it being benign.\\

\noindent\textbf{OriginMatch.} Conflicts may arise due to incorrect configurations of the MaxLength parameter in ROAs \cite{gilad2017maxlength}. When such conflicts occur, we set this feature value to 1 because the source in the BGP announcement matches the origin in the ROA, indicating that the announcement originates from the IP address prefix's authorized owner.

We retrieve validated prefix origin data (i.e., ROAs) from the RPKI repositories to determine the outcome of ROV and the feature \textit{OriginMatch}.

\noindent\textbf{PC.} Conflicting origins in RPKI validation may sometimes involve a provider and its customers.
For instance, consider a provider AS3215 with prefix \texttt{193.2.0.0/15} that delegates some of its IP prefixes to its two customers: AS1272 \texttt{193.2.35.0/24} and AS8361 \texttt{193.2.155.0/24}. However, the provider AS3215 issues a ROA for its prefix \texttt{193.2.0.0/15} for origin AS3215, but the customers do not issue ROAs for their prefixes. When the customers announce routes for the prefixes, conflicts occur. 
Such conflicts are typically benign as the customer AS is unlikely to hijack its providers.
Thus, if a provider and customer relationship is found between the two conflicting origins, the feature value is designated as 1.

We retrieve the information about AS relationships from CAIDA \cite{CAIDA} to compute the feature \textit{PC}.

\noindent\textbf{MOAS.} When multiple ASes advertise routes for a given prefix, but only one AS has created the ROA for it, this scenario can result in a multiple origin AS (MOAS)
conflict \cite{chin2007characteristics,zhao2001analysis}. MOAS conflicts can arise for reasons such as multi-homing, multinational companies, or organizations with multiple AS numbers \cite{chin2007characteristics}, which are generally considered benign.
If the two conflicting origins of a route exhibit MOAS conflicts (e.g., they are from the same organization), we assign 1 to this feature.

In this work, we focus solely on AS organization information from CAIDA to compute the \textit{MOAS} feature. Additional data that could be used in computing \textit{MOAS} will be explored in future work.

\noindent\textbf{Parent.} In some cases, a parent organization splits a prefix into a set of more specific prefixes, assigning them to its child organizations with their own AS numbers. Assume separate ROAs are created for both the parent and the child prefixes. When the parent AS announces the prefix on behalf of its child, a conflict arises. If we detect that the originating AS has another prefix in ROAs, which is less specific than the declared prefix, the feature value is set as 1.

The computation for the feature \textit{Parent} relies on the ROAs from the RPKI repositories.

\begin{table}[t!]
  \centering
  \renewcommand{\arraystretch}{0.8}
  \fontsize{5}{5}\selectfont
  \begin {tabular}{c|c|c}
  \toprule
  \textbf{Feature} & \textbf{Definition} & \textbf{Source} \\\midrule
    \begin{tabular}{@{}c@{}}
                OriginMatch \\
                   PC \\
                   MOAS \\
                   Parent \\
                   Depen \\
                   AltSources\\
                   ASdist \\
                 \end{tabular} & \begin{tabular}{@{}c@{}}
                   
                   The origin matches but the MaxLength mismatches. \\
                   A relation of provider and customer exists between two origins.\\
                   Two origins belong to MOAS.\\
                   A parent prefix is already in ROAs.\\
                   Two origins have a strong dependency. \\
                   Other trusted sources validate the route as valid.\\
                   The geographical distance between two origins. \\
                   \end{tabular} & \begin{tabular}{@{}c@{}}
                   
                   RPKI \\
                   CAIDA\\
                   CAIDA\\
                   RPKI\\
                   IHR\\
                   IRRs\\
                   CAIDA, GeoIP\\
                   \end{tabular} \\\bottomrule
   \end{tabular}
\caption[Features along with their definitions.]{\small Features along with their definitions. "Source" refers to the data source for feature computation.}
    \label{tab:features}
    \vspace{-15pt}
\end{table}

\noindent\textbf{Depen.}
When customers acquire anti-DDoS services from a provider, they may authorize the provider to announce routes for their IP prefixes. The provider then directs the traffic destined for these prefixes back to the customer. In this arrangement, discrepancies between the published ROAs and the actual prefix announcements can arise. For instance, the provider may publish ROAs for prefixes that are owned by the customer, while announcing these prefixes with the customer’s ASN appended to the end of the AS path. This can result in benign ROA conflicts.

Note that, the service provider and the service customer may not have a direct BGP relationship (e.g., PC). We propose a new relationship between two conflicting origins, referred to as \textit{Depen}, which represents the interdependency between the two origins. For instance, in the scenario mentioned above, the customer demonstrates a dependency on the service provider for accessing the Internet.

The Internet Health Report (IHR) \cite{IHR} introduces the AS hegemony metric to estimate AS interdependencies using global routing information \cite{fontugne2017hegemony,fontugne2018thin}. AS hegemony comprises two types: local and global. Local hegemony measures the dependency between any two ASes, while global hegemony indicates the centrality of an AS on the Internet. 

We leverage local hegemony to compute the \textit{Depen} feature. In cases of bidirectional dependencies between two origins, we assign the larger local hegemony value as the \textit{Depen} value.
The AS hegemony metric ranges from 0 to 1, with greater values indicating stronger dependencies.
When two origins exhibit a greater AS hegemony value, it signifies a closer connection between them. This, in turn, suggests that the route is likely benign.

\noindent\textbf{AltSources.} This feature is utilized to check if the RPKI-invalid route has a valid validation outcome when using other reliable sources. When it does, it implies a potential (albeit undetermined) association between two conflicting origins. 

The IRRs \cite{IRR} are distributed databases managed by RIRs like APNIC, RIPE NCC, or Internet Service Providers (ISPs) such as Level 3. They hold valuable routing information, including AS numbers and their corresponding IP addresses.
Therefore, the IRRs can be considered to be alternative sources for route validation. 
To calculate the feature \textit{AltSources}, we use the \textit{whois} tool to extract IRR entries (i.e., AS numbers and IP prefixes).
Once the RPKI-invalid route is verified as valid based on IRR data, its feature value is assigned as 1.

\noindent\textbf{ASdist.} This feature is utilized to assess the degree of geographical proximity between two conflicting origins. Intuitively, unlike benign conflicts, origins involved in a prefix hijack are often situated at a greater distance from each other, as there are typically no direct connections between them.
Given two origins, $AS_{x}$ and $AS_{y}$, we denote $g_{x}$ ($\phi_x$, $\lambda_x$) and $g_{y}$ ($\phi_y$, $\lambda_y$) as the geographical locations of the two ASes, with $\phi$ and $\lambda$ being latitude and longitude coordinates, respectively.
Then we compute great circle distance (which is commonly used to calculate the shortest path between two locations on Earth) between $g_{x}$ and $g_{y}$, denoted as $d$.

We utilize MAXMIND's GeoIP databases to obtain latitude and longitude information for IP prefixes within origin ASes. Since each AS may contain prefixes with different geographic locations, one method is to compute the geometric median of these locations to represent the AS’s location. However, this can be computationally intensive. To simplify, we use the location of the announced prefix to represent the location of the RPKI origin AS and calculate the great circle distance between this location and each prefix (if the number of prefixes exceeds 500, we randomly select 500 prefixes) within the origin AS in BGP announcements. The median of these distances is then used as the feature value $d$.

\begin{figure}[t!]
    \centering
    \includegraphics[width=0.7\linewidth]{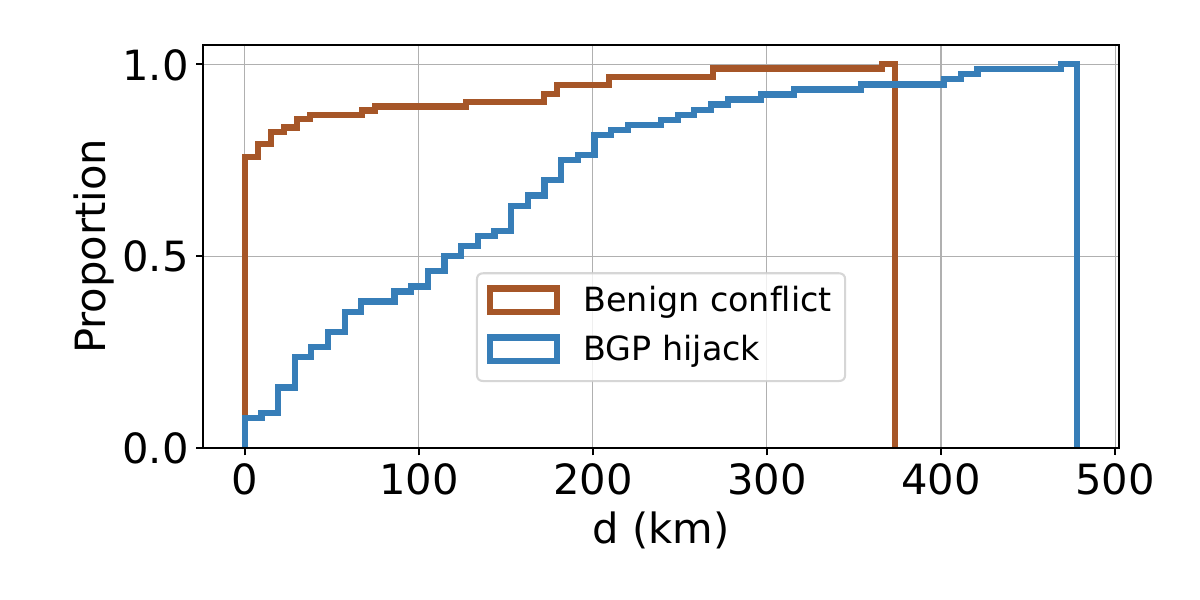}
    \vspace{-15pt}
    \caption{\small CDFs of $d$.}
    \label{fig:d_cdf}
\end{figure}

Figure \ref{fig:d_cdf} illustrates the cumulative distribution functions (CDFs) of distances ($d$) we analyzed in 100 benign conflicts and 100 prefix hijacks randomly selected from our ground truth dataset (see details about the ground truth data in the following paragraphs).
The plot shows that benign conflicts mostly (90\%) exhibit a smaller distance value (less than 50km) compared to hijacks, which aligns with our previous assumption.

Compared with other features that fall into [0, 1], $d$ exhibits a larger upper bound. The larger values may cause bias in the classification model (e.g., in cases of using SVM). Hence, we scale this feature before training.
In our experiments, we discovered that scaling $d$ into the range of [0, 1] using a commonly used method like minimum-maximum scaling may result in a number of false negatives, i.e., misidentifying hijacks as benign conflicts. These mis-recognitions often occur when hijacks entail two conflicting origins that are in close proximity.
To address this, we apply the arctangent function to $d$, as given by the formula: {$ASdist = \frac{2}{\Pi}arctan(d)$}, resulting in a new distance metric $ASdist$ called AS distance.
Compared to $d$, $ASdist$ is expected to be more effective at distinguishing between benign conflicts and hijacks in close proximity (see more details about this method in Appendix \ref{app:feature_scale}).
\vspace{-5pt}
\subsubsection{Utilization of Default Values}
We expect benign conflicts to show a closer relationship between the two origins, characterized by more connections and shorter distances between them.
The default value for the features is 0, except for $ASdist$, which is set to 1, indicating no relations between conflicting origins. These default settings make instances more likely to be categorized as BGP hijacks, which helps uphold the effectiveness of ROV. Feature computation often utilizes reliable data sources, which rarely become unavailable.
In cases where required information is missing, default values are used, which may result in benign conflicts being misidentified as hijacks. This is acceptable because, although our primary goal is to detect benign conflicts, it is premised on maintaining ROV's hijack prevention.
\vspace{-10pt}
\subsubsection{ML Identification}
While using these features as rules to distinguish between benign conflicts and real hijacks could be effective, such an approach may overlook the overall relationship strength (i.e., tightness) between conflicting origins in a given BGP route. In contrast, ML allows these features to work collectively, capturing nuances that humans might miss, potentially reducing the impact of inaccuracies in certain data sources. For instance, a \textit{PC} relationship might appear in some hijacks, possibly due to errors in CAIDA, while a rules-based approach could falsely classify this as a benign conflict. Moreover, features like \textit{ASdist} require predefined thresholds for classification. Heuristic-based methods, which rely on manual thresholds, may not adapt well to changing conditions. On the other hand, ML methods automatically identify patterns in input features, often resulting in better classification performance. Therefore, in this work, we opt to use ML methodologies for classification.

\vspace{-5pt}
\subsection{Implementation, Evaluation, and Analysis} \label{subsec:groundtruth}

\subsubsection{Data Collection} We first collect a real-world ground truth dataset of ROA-violating routes, including benign conflicts and BGP hijacks, for training and validating the classifier.

\noindent\textbf{Benign conflicts.} According to previous studies \cite{vervier2015mind, wijchers2014bgp, mahajan2002understanding}, hijacked BGP announcements are usually short-lived (e.g., lasting only a few hours or days). In other words, long-lived announcements are generally considered benign. Using this basic concept, we collect BGP routes with ROA violations.

We first utilized Routinator \cite{Routinator}, a widely recognized and reliable open-source RPKI relying party validator (ROV), to validate BGP announcements that occurred between May 30, 2022, and June 30, 2022, collected from the RIB data of RouteViews.
We focused specifically on routes that remained stable and persistent throughout the period, ensuring their benign nature. Note that we may encounter multiple announcements for the same prefix, originating from the same AS but with different AS paths. In such cases, we retained only a single route. The validation process identified 340,485 RPKI-valid announcements and 9,223 RPKI-invalid ones.
The invalid announcements are considered to be routes with benign ROA conflicts, originating from 1,377 different ASes.
 
\noindent\textbf{BGP hijacks.} BGPmon \cite{BGPMon} is a reliable source that provides various BGP events from different ASes at different times \cite{van2019analysing}. In addition, BGPmon offers a specific instance for each event, ensuring the diversity of the ground truth. We collected BGP hijacking instances from BGPmon, which occurred between December 2021 and September 2022.
We then cleaned the data by filtering out cases that either conformed to ROAs or were not covered by ROAs, resulting in 415 hijacks initiated by 246 ASes.

\noindent\textbf{Overview of Ground truth data.} Consequently, we collected 9,223 benign conflicts and 415 hijacks. Given that there is a relatively large imbalance (22:1) between the two classes, we randomly select 2,000 instances from the benign conflict dataset. This reduction helps alleviate the imbalance, making it easier to handle while preserving the diversity of benign conflicts.
Then, we increase the number of instances in the minority class to 2,000 using random oversampling. This involves randomly copying instances from the minority class, which is often well-suited for smaller imbalances, resulting in a ground truth set including 2,000 benign conflicts and 2,000 BGP hijacks. The remaining 7,223 will be used as new or unseen data for evaluation in Section \ref{subsec:holdout}.

\noindent\textbf{Limitations.}
Collecting ground truth data from the real world is often challenging because network operators typically keep their data private and confidential. To address this issue, we leverage the long-duration nature of benign routes to identify and extract benign conflicts from RPKI-invalid BGP routes. While this approach helps ensure that the collected benign conflicts are likely to be genuinely benign, it may miss short-lived conflicts, such as those lasting only a few hours or days. Consequently, the absence of transient benign conflicts in the ground truth data could impact data diversity and potentially bias model performance. On the other hand, although BGPmon is highly regarded as a reliable source, it may still encounter issues with false positives. We filtered out instances mistakenly identified as hijacks to ensure the cleanliness of the collected data. However, due to the lack of transparency in BGPmon’s operational process, we cannot guarantee that the remaining BGP hijacks are genuinely malicious. Moreover, while random oversampling can help address class imbalance, it might introduce certain biases (e.g., distorting data distribution). Future work will explore more effective strategies to construct ground truth data.
\vspace{-5pt}
\subsubsection{Cross-validation on Ground Truth Data}
The classifiers based on the ML models mentioned above are implemented using the scikit-learn (sklearn) library in Python 3.8 and validated on the ground truth data.
The data is split into training and testing sets using an 8:2 ratio. 10-fold cross-validation is conducted to assess the model performance, using measures including macro-averaging precision, recall, and F1-score.
Additionally, grid search is employed for hyperparameter optimization, focusing on the primary parameters of the models. Since NB has relatively few tunable hyperparameters, the default settings are employed.
The grid search settings for the other four models (DT, SVM, KNN, and RF) are presented in Table \ref{tab:identi_parameters} in Appendix \ref{app:grid_search_settings}.

Table \ref{tab:lov_classifiers_res} displays the optimal hyperparameters searched for the five binary classification models. In addition, the evaluation results on the ground truth data set are also presented in the table. It can be observed that, the DT, KNN, and RF models achieve higher accuracy than the other two models, with precision, recall, and F1-score all around 99\%.

\begin{table}[t!]
  \centering
  \fontsize{7}{7}\selectfont
  \renewcommand\arraystretch{1}{
\begin{tabular}{ccccccc}\toprule
 & \textbf{DT} & \textbf{SVM}  & \textbf{KNN} & \textbf{RF} & \textbf{NB}\\\midrule
 \textbf{Precision} & 98.7\% & 98.1\% &  99\% & 98.8\% & 97.2\% \\
 \textbf{Recall} & 98.7\% & 98.1\% & 99\% & 98.8\% & 97.1\% \\
 \textbf{F1-score}  & 98.7\%  & 98.1\% & 99\%  & 98.8\% & 97.1\% \\ \midrule
 \textbf{New (``Benign'')} & 96.1\% & 94.4\% & 94.8\% & 94.6\%  & 94.2\% \\
 \textbf{New (``Hijack'')} & 98.1\% & $\sim$100\% (1)  & $\sim$100\% (4) & $\sim$100\% (1) & 96.9\%  \\
\bottomrule
\end{tabular}
}
 \caption[Evaluation results of five ML models.]{\small 
 Evaluation of macro-averaging precision, recall, F1-score obtained on the ground truth, accuracy when tested on the new dataset (``Benign'' and ``Hijack'' refer to benign conflict instances and BGP hijack instances). We note the number of detection errors in the brackets in cases of $\sim$100\%.}
    \label{tab:lov_classifiers_res}
    \vspace{-15pt}
\end{table}
\vspace{-10pt}
\subsubsection{Evaluation on New Data} \label{subsec:holdout}
Subsequently, we collect a new set for evaluating the effectiveness of the five ML models we trained above, specifically assessing how well the model generalizes to new, unseen data.
Recall that, we only selected 2000 instances as ground truth data. The remaining 7,223 benign conflicts are used subsequently as new data for the evaluation.
Additionally, we obtain new hijacks from another well-known data source. We describe the collection procedure in detail below.

Qrator Labs \cite{QratorLab} provides a BGP monitoring service called Radar that assists researchers and network operators in detecting network anomalies. In contrast to BGPmon, Radar focuses more on malicious events that have a significant impact on Internet services at the global routing level. To create a new data set as the holdout data, we collected 8 BGP hijacking events from Radar, which happened between the years 2020 and 2022, as listed in Table \ref{tab:bgp_incidents_res}. BGPStream also provides access to historical BGP data. We applied it to collect BGP announcements associated with the aforementioned incidents, through three RouteViews collectors: \textit{amsix}, \textit{wide}, and \textit{chicago}, located in Amsterdam, Japan, and the USA, respectively.
For each incident, we collected routes originating from the hijacking AS over the time period of the incident and removed announcements with compliant ROAs or unknown prefixes to ROAs.
This resulted in a collection of 10,068 hijacks.
Note that we retained only one instance of a hijack when multiple hijacks involved the same hijacking AS and the same victim AS (i.e., the RPKI origin), thereby reducing redundancy in hijack identification.

The evaluation results on the new set are also presented in Table \ref{tab:lov_classifiers_res}.
Despite achieving a higher accuracy of 96\% in identifying benign conflicts, the DT model exhibits a relatively low performance in detecting hijacks, with an accuracy of around 98\%. In contrast, the SVM, KNN, and RF models achieve $\sim$100\% accuracy in hijack detection, meeting our requirements for the ML-classifier.
While the KNN model slightly outperforms the RF in benign conflict identification, the RF exhibits fewer detection errors, with only one instance being misidentified.

Our evaluation results show that the RF model exhibits better performance overall. Consequently, we select the RF-based classifier to differentiate benign conflicts from BGP hijacks.

 \begin{table}[t!]

  \centering
  \fontsize{5}{5}\selectfont 
  \renewcommand\arraystretch{0.9}{
  \begin {tabular}{ccccccc}
  \toprule
    \textbf{Time} & \textbf{Hijacker ASN} & \textbf{Organization} & \textbf{VP (\#)} & \textbf{Duration} & \textbf{Hijacks (\#)} & \textbf{Acc} \\
    \midrule
   2020-04-01 & AS12389 & Rostelecom & 8870 & 1 hour & 7396 & 100\% \\
   2020-09-29 & AS1221 & Telstra & 472 & 3 hours & 58 & 100\% \\
   2021-04-16 & AS55410 & Vodafone Idea & 37,739 & 1.5 hours & 1321 & 100\% \\
   2021-05-18 & AS48467 & PRANET  & 454 & 5.7 hours & 99 & 99\% \\
  2021-10-13 & AS212046 & MEZON & 1029 & 1 hour & 449 & 100\% \\
   2021-10-25 & AS212046 & MEZON & 3786 & 0.5 hours & 24 & 100\%  \\ 
   2022-01-31 & AS18978 & ENZUINC & 743 & 0.7 hours & 274 & 100\%\\
   2022-05-31 & AS38744 & AONB & 461 & 0.4 hours & 447 & 100\% \\
   \bottomrule
   \end{tabular}
   }
   \caption[Statistics of BGP hijacks collected.]{\small{BGP hijacks we collected, and evaluation with RF-based classifier for each event. VP: victim prefixes, Acc: accuracy.}}
\label{tab:bgp_incidents_res}
\vspace{-15pt}
\end{table}

\begin{figure}[b!]
\centerline{\includegraphics[width=0.6\linewidth]{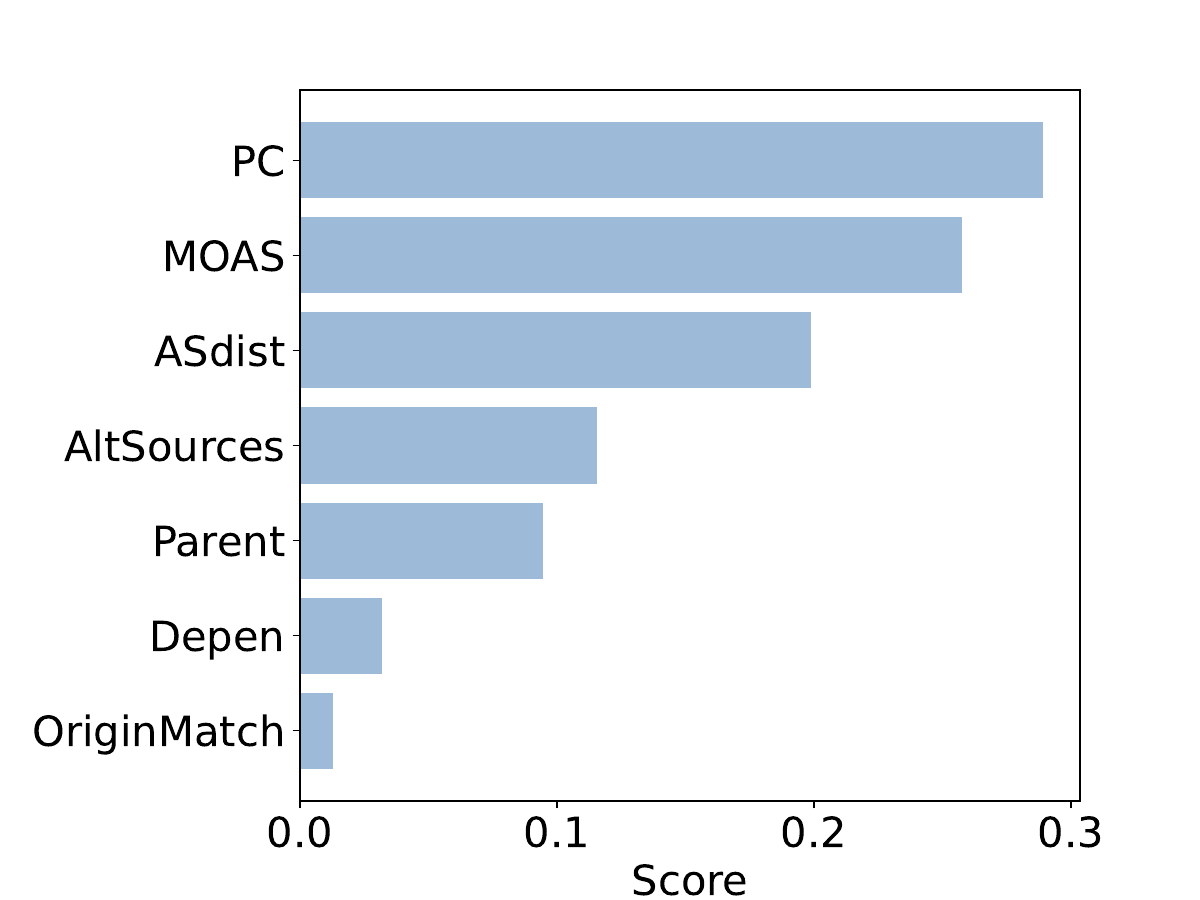}}
\vspace{-10pt}
\caption{\small{Feature importance.}}
\label{fig:lov_feature_importance}
\end{figure}

\vspace{-8pt}
\subsubsection{Feature Importance} \label{subsubsec:lov_feature_importance}
We assess the importance of individual features in the RF classifier's decision-making. Figure \ref{fig:lov_feature_importance} shows the feature importance scores for the seven features used. The features $PC$, $MOAS$, $ASdist$, $AltSources$, and $Parent$ have the most significant influence, each with a score greater than 0.1 and collectively contributing over 0.95. The frequent absence of local AS hegemony values possibly reduces the impact of $Depen$, with around 35\% of the routes in our ground truth set lacking this data.
Routes with the same origin satisfy $MOAS$ but not vice versa, reducing the importance of $OriginMatch$ and increasing that of $MOAS$. Conflicting origins with strong AS dependency, potentially involving a provider-customer relationship, may amplify the impact of $PC$.

\vspace{-8pt}
\subsubsection{Errors Analysis}
Next, we analyze the detection errors of the RF classifier.
In some benign conflicts, no relationship between the conflicting origins was detected. This may stem from a lack of relevant information in the data sources, particularly for newly emerging ASes, which led to the use of default values and subsequent misidentification.
Additionally, benign conflicts involving human typo errors in ROA issuance—such as AS4433 being erroneously written as AS443, often manifest as real hijacks, possibly leading to detection failures.

In contrast, the undetected hijack was likely caused by a smaller AS distance value between the two origins. To assess the success rate of such attacks, we crafted 5,000 prefix hijack instances originating from 5,000 randomly selected ASes.
BGP neighbors (such as providers, customers, and peers) can be relatively close geographically. The targets of these hijacks were IP addresses from neighboring networks with a peering relationship with the hijacking AS.
We excluded those neighbor networks with provider or customer relationships to avoid potential benign origin errors. 1.6\% of these hijacks were misclassified as benign by the classifier, indicating that the likelihood of adversaries launching such attacks is low.

\vspace{-5pt}
\subsubsection{Robustness to Adversarial Attacks} \label{subsubsec:robustness}
As discussed above, adversaries might design BGP hijacking attacks to evade detection by the ML-classifier, though such attacks have a low success rate.
Recall that benign conflicts identified by the ML-classifier are not immediately added to the whitelist. Instead, they are quarantined for further review, which includes examining the tightness of the conflicting origins associated with a route and monitoring their longevity or activity level. Therefore, even if such hijacks succeed in bypassing the classifier's detection, the rigorous review process makes it less likely that these hijacks will be whitelisted, as they typically exhibit lower tightness and shorter duration.
The quarantine mechanism can effectively prevent adversarial attacks from compromising the whitelist, ensuring its reliability.
Please refer to Section \ref{subsec:quarantine} for more details.

Additionally, adversaries may also manipulate the data used for calculating features to avoid identification. However, the public data resources we have meticulously selected are characterized by high reliability, trustworthiness, and robust maintenance. These qualities make it challenging for attackers to tamper with the data, thereby reducing the likelihood of such attacks being attempted.
Regarding the other two common types of adversarial attacks, namely poisoning attacks and model extraction attacks, they pose less of a threat to the effectiveness of \lov. Attackers would find it impossible to modify the training data of classification models because the data is kept private and secure to prevent unauthorized access. Additionally, the training data used by \lov\ does not contain any sensitive or confidential information that an attacker would be interested in extracting.
\vspace{-8pt}
\subsubsection{Model Generalization} \label{subsubsec:model_general}
Our feature set focuses solely on capturing the relationships between two conflicting origins rather than the behavior patterns underlying the routes.
This can mitigate the impact of the evolving behavior of anomalous routes over time on identification, thereby minimizing the model generalization issues to newly emerging data.
Additionally, the feature values may vary when updated public data sources are used, potentially biasing the classification. However, this variation is not due to the features themselves, but rather to inaccuracies in the data sources.

Appendix \ref{app:ml_challenges} discusses additional challenges faced during the development and deployment of the ML-based classifier.

\vspace{-5pt}
\section{Post-Analyzer} \label{subsec:lov_post_analyzer}


\subsection{Anomalies in AS Visibility}
The proposed post-analyzer approach relies on the bursty property of BGP announcements during a BGP incident, in which groups of routing messages are sent in short intervals.
We leverage this property to detect whether suspicious announcements identified by the RF classifier are a result of a BGP event. Prior studies such as \cite{moriano2021using, deshpande2009online} counted the total number of announcements over a certain period to identify routing anomalies. However, these methods involve processing large volumes of BGP routes, making them resource-intensive and time-consuming.
In contrast, we employ global AS visibility, which refers to the extent to which an AS is visible on BGP paths in global routing tables.
In general, the AS visibility of reliable networks tends to remain stable. However, during a hijack event, the offending AS illegitimately originates routes for a substantial number of prefixes that are not under its ownership, causing its global visibility on the Internet to significantly increase. We leverage this change in global AS visibility as a detection signature for hijack events.
\begin{figure}[t!]
\centerline{\includegraphics[width=0.8\linewidth]{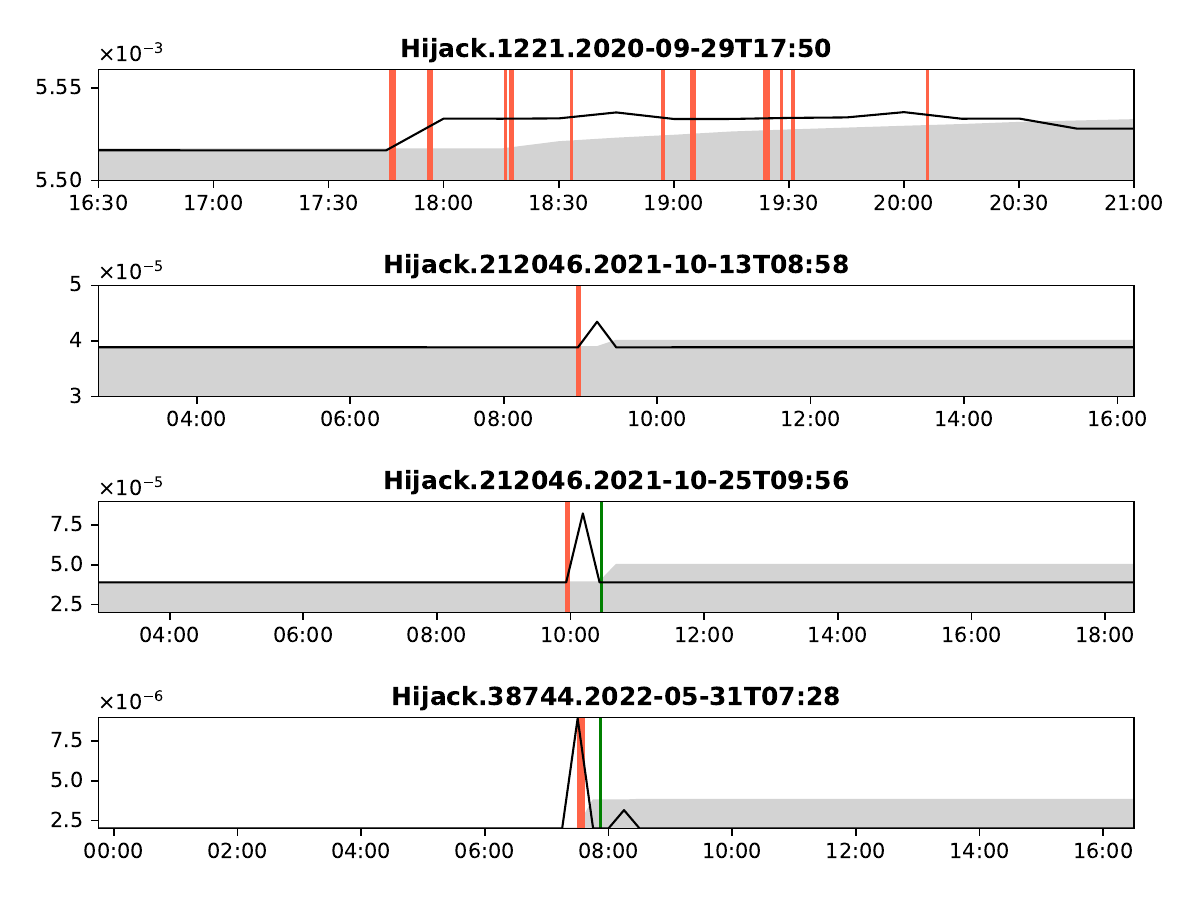}}
\vspace{-15pt}
\caption[Changes in global visibility of malicious ASes.]{\small{Changes in global visibility of malicious ASes before, during and after the BGP incidents and evaluation with the post-analyzer. Red vertical lines indicate outlier detections.}}
\label{fig:post_analysis}
\end{figure}

To calculate the global visibility of an AS at a particular time, we employ global AS hegemony provided by IHR.
Unlike local AS hegemony, global hegemony reflects the centrality of an AS on the path, indicating the likelihood that the AS appears on BGP paths in the routing table. Therefore, this metric can also serve as an indicator of AS visibility.
We detect the anomalous increase in AS visibility by monitoring for a surge in global hegemony values.
The use of AS hegemony offers a more direct and efficient approach to detecting abnormal BGP behavior (e.g., hijacking), compared to previous methods.
The IHR service provides AS hegemony values in real-time.
\vspace{-5pt}
\subsection{Z-test Anomaly Detection}
We then leverage a Z-test to detect anomalies in the hegemony values of hijacking ASes.
For each suspicious BGP announcement identified by the classifier, the post-analyzer collects the most recent historical hegemony values of the perpetrator AS preceding the announcement time.
These values are modeled as a variant with Gaussian distribution, denoted as $X$. We then calculate Z = $\frac{X-\mu}{\sigma}$, which follows the standard normal distribution of $N(0,1)$, with $\mu$ and $\sigma$ being the mean and standard deviation of $X$. Next, we evaluate the p-value of a hegemony value observed after the announcement to determine if it is statistically significant.

We collect 50 historical hegemony values, which provides sufficient data to accurately capture their distribution while minimizing the impact of older, potentially less relevant data, on the p-value calculation.
We specify a null hypothesis that there is no obvious difference between the tested hegemony value and historical data at a significant level of 0.05.
Since the perpetrator AS has an increasing hegemony value during the event, we perform a right-tailed test to detect anomalies. If the p-value is greater than 95\%, the null hypothesis is rejected, indicating a potential hijack incident.
We also monitor whether the hegemony value returns to its normal level after the event. If the hegemony value remains consistent for a long period (e.g., 24 hours), we consider the possibility of a new routing policy in place that is causing a different level of AS visibility.
While the Z-test is straightforward, it is sufficiently reliable for detecting anomalies in AS hegemony values.

\vspace{-10pt}
\subsection{Evaluation and Analysis}
We then evaluate the post-analyzer.
To simplify this process, we use 8 BGP incidents collected in Section \ref{subsec:holdout} (refer to Table \ref{tab:bgp_incidents_res}) for the evaluation.
For a given incident, we extract timestamps from each of the route announcements during the event and verify that the hegemony value at the announcement time is abnormal.

We detected anomalies in global AS visibility for four incidents, as illustrated in Figure \ref{fig:post_analysis}.
We mark each incident in the format: AttackType.AttackerASN.TimeOfOccurrence. 
The gray area refers to the safe region of hegemony values, and the upper bound of this region means the lower threshold of identifying outliers.
Red vertical lines indicate that abnormal AS hegemony values were detected at corresponding BGP announcement times, while green vertical lines mean that the post-analyzer has not detected anomalies.
As can be seen, the post-analyzer failed to identify anomalies at the later stage of two events: ``Hijack.212046.2021-10-25T09:56'' and ``Hijack.38744.2022-05-31T07:28''.
The reason for this failure was likely that the incidents were immediately detected, and the corresponding BGP announcements were filtered to prevent their propagation on the Internet, resulting in a decrease in the perpetrators' global visibility.

In addition, for the event "Hijack.48467.2021-05-18T09:00", increased hegemony values were observed starting from 7:00. However, this rise in hegemony values caused an increase in the outlier detection threshold at 9:00 (the event occurrence time), causing the post-analyzer to fail to verify this event.
For the remaining three BGP events, no corresponding hegemony values were available during the event periods, which prevented the post-analyzer from tracking changes in the perpetrator ASes' global visibility.
We attribute these failures to potential issues in the IHR's hegemony computation mechanism.


Note that, even if the post-analyzer does not confirm certain events, these limitations have minimal impact on the reliability of the whitelist.
Unverified routes will be placed in quarantine for additional examination.
One strategy for whitelisting involves continuously monitoring the behavior of these routes, such as their longevity (see more details in Section \ref{subsec:quarantine}).
Given the typically transient nature of hijacking incidents, routes associated with hijacking, that are mistakenly placed in quarantine, are less likely to meet the criteria for inclusion in the whitelist. 

\section{Quarantine} \label{subsec:quarantine}
As mentioned in Section \ref{sec:lov_overview}, we employ a quarantine mechanism for further inspection to ensure the quality of the whitelist.
We employ different strategies to assess quarantined routes for reliable whitelisting, tailoring these strategies to the priority level of the routes while also considering whitelist quantity and efficiency. 

Benign conflicts identified by the classifier are given high priority for whitelisting, as the classifier's false negative rate (misclassifying hijacks as benign conflicts) is expected to be very low, given its design goal. Nevertheless, reviewing these routes will be beneficial for countering adversarial attacks, specifically evasion attacks, as described in Section \ref{subsubsec:robustness}.

Recall that the tighter the relationship between two origins involved in a BGP route, the more likely the route is to be benign.
For these routes, we thus examine the overall tightness level between the two origins. We evaluate the tightness between the two origins by defining a formula: {$T = w_{0}*OriginMatch + w_{1}*PC + w_{2}*MOAS + w_{3}*Parent + w_{4}*Depen + w{5}*AltSources - w_{6}*ASdist$}, where $T$ refers to the overall tightness value, and $w_{0}, w_{1}, w_{2}, w_{3}, w_{4}, w_{5}, w_{6}$ represents the positive weight values for the seven features used.
This formula is designed to be straightforward in computation and is based on the intuitive concept that two origins with more relations and smaller distance values (such as $PC = 1$ and $Depen = 1$ and $ASdist = 0.01$) exhibit greater tightness.
We specify the $w_{0}, w_{1}, w_{2}, w_{3}, w_{4}, w_{5}, w_{6}$ values using the feature importance scores computed in Section \ref{subsubsec:lov_feature_importance}.
The first strategy we utilize is to whitelist routes that exhibit a $T$ value greater than $T_{thr}$ (the threshold value to be determined in our measurements).
This approach ensures a sufficient number of routes in the whitelist while maintaining high efficiency due to its simplicity.

Next, for routes excluded by the first strategy and those not confirmed by the post-analyzer, we carefully monitor their behavior—including their longevity and activity levels—drawing inspiration from the PGBGP tool (as mentioned in Section \ref{sec:relatedwork}).
A longer monitoring period increases confidence that whitelisted routes are indeed benign, given that benign routes typically persist longer.
However, extending this period decreases the efficiency of the whitelisting process.
Based on our empirical observations, we set the quarantine period to 14 days, sufficiently long to determine whether a route is benign while maintaining reasonable efficiency. Additionally, to guarantee the relevance of the whitelist, our primary focus is on whitelisting actively used routes. \lov\ employs two criteria for this whitelisting. First, the route must appear on at least two days within a week, ensuring adequate frequency of occurrence. Second, the most recent appearance of the route should be no more than a week before the end of quarantine, guaranteeing that the route is likely still active.
Note that this behavior monitoring strategy is more rigorous (i.e., involving long-term monitoring and focusing solely on routes in active use) compared to the first one, as it addresses routes that are considered to have a relatively lower likelihood of being benign.
This approach whitelists fewer routes and exhibits lower efficiency due to the stringent rules.

Moreover, we introduce human intervention as a strategy for whitelisting. As mentioned earlier, evidence obtained from the analysis of hijacking events through email surveys assists security analysts in implementing this intervention. For instance, if email surveys indicate that routes of interest, which have conflicting origins with ROAs, are genuinely benign and not due to correctable errors, we manually add these routes to the whitelist if they are not already included. Conversely, if it is confirmed that the conflicting origins are the result of hijacking, we ensure these routes do not pollute the whitelist. Human intervention enhances the effectiveness and reliability of the whitelist.

\vspace{-10pt}
\section{Real-world Measurements} \label{sec:measure}
In this section, we applied \lov\ to measure BGP updates occurring between October 1, 2022, and March 31, 2023, with daily tests on unique routes.
We validated a total of 117,293,602 BGP announcements, containing 68,507,964 routes with prefixes that were unknown to ROAs, 47,737,942 that were RPKI-valid, and 1,047,696 that were RPKI-invalid. We excluded hijacks originating from a bogon AS number from analysis as they are easily identified.

\vspace{-5pt}
\subsection{Results and Analysis}

\subsubsection{Identified Benign Conflicts By the ML-Classifier}
\begin{figure}[t!]
\centerline{\includegraphics[width=0.8\linewidth]{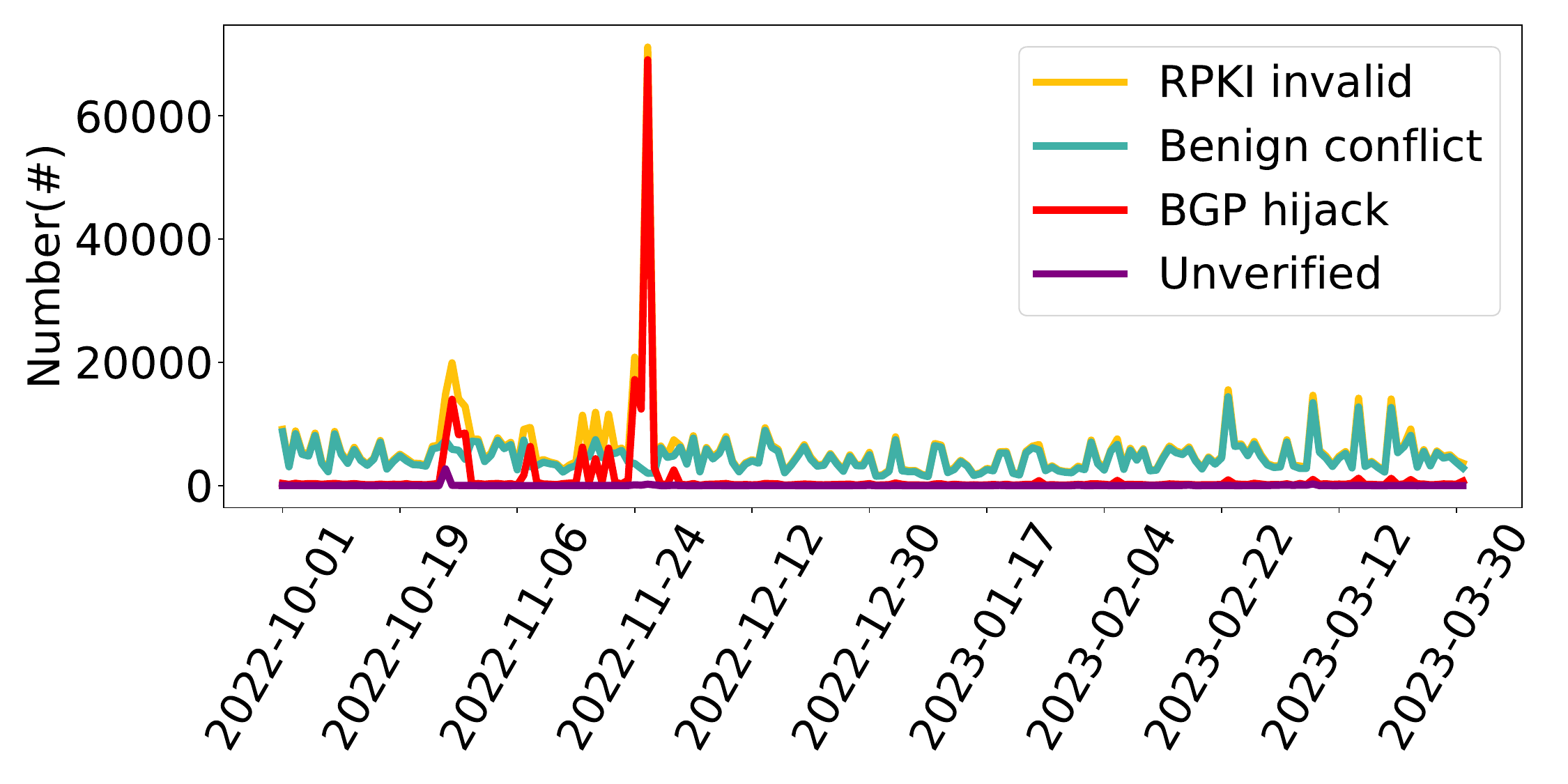}}
\vspace{-15pt}
\caption[The daily number of RPKI-invalid routes, benign conflicts, and hijacks.]{\small{The daily number of RPKI-invalid routes detected by ROV, benign conflicts, hijacks identified by the RF classifier, and hijacks unverified by the post-analyzer throughout the measurement period.}}
\label{fig:invalid_bc}
\end{figure}

Figure \ref{fig:invalid_bc} shows the RPKI-invalid routes identified by ROV, as well as the benign conflicts and BGP hijacks detected by the RF classifier throughout the measurement period.
We detected a total of 183,988 objects (each object containing a unique pair of the originating AS and the announced prefix) resulting in RPKI status invalid, covering 102,427 prefixes and 30,775 ASes. The number of RPKI-invalid routes occurred with an average of 5,757 instances per day. With the RF classifier, we identified 63,458 benign conflicts, covering 62,569 prefixes from 5,110 ASes, with an average daily occurrence of 4,549.
On a daily average, approximately 79\% of RPKI-invalid routes are likely attributed to benign conflicts with ROAs.

The occurrences and frequencies of these benign conflicts are displayed in Figure \ref{fig:occur_fre_cdf} in Appendix \ref{app:cdf_benign_conflicts}. An ``occurrence'' refers to the appearance of a particular benign conflict during a day. 
 We observed 33,191 (52\%) benign conflicts that appeared on at least two days during the measurement period, and 80\% of them had an average occurrence frequency of less than 14 days.
 This observation is in line with our expectation that benign conflict tends to be persistent and used actively.
\vspace{-5pt}
\subsubsection{Identified Hijacks By the ML-Classifier}
Additionally, 119,815 hijacks were detected during the measurement period, with a mean of 1,166 occurrences per day.
As shown in the figure, massive hijacks were identified during four periods: between 2022-10-26 and 2022-10-29, between 2022-11-07 and 2022-11-08, between 2022-11-16 to 2022-11-21, and between 2022-11-23 to 2022-11-30.

We conducted a specific analysis of the anomalous routes identified during these four periods. We detail the statistics in Table \ref{tab:specific_days_res} in Appendix \ref{app:statistics}.
We discovered that during the four periods, 14,170, 6,720, 7,388, and 84,157 routes resulting in RPKI invalid status were exported through AS212483, which was one of 41 peers that fed BGP data to a RouteViews collector we used, accounting for 58\%, 45\%, 39\%, and 86\% of the invalid routes, respectively. As shown in the table, most of these routes were identified as hijacks by our classifier, accounting for around 99\%, 100\%, 99\%, and 99.7\%, respectively. We believe that it was not a coincidence that AS212483 was linked to the majority of anomalous routes that occurred during these four periods.

To investigate the underlying causes for these suspected routes, we reached out to the network operators of the ASes potentially responsible for the anomalous routes.
To expedite this process, we randomly selected 500 routes initiated by different ASes and obtained the email addresses of respective network operators using the \textit{whois} tool.
We described the incident that affected the AS in question and requested possible explanations in our email. We received responses from 10 network operators, all of whom denied having announced the anomalous routes. Of particular note was the response from the network operators of AS50058, who informed us that AS212483 had a history of feeding incorrect BGP data to route collectors.
This strongly suggested that AS212483 was the likely source of the large number of anomalous routes observed during the measurement period.

We also attempted to contact the operators of AS212483 but received no explanation for the issues we observed. Our hypothesis is that the anomalous routes may have been caused by misconfigurations or technical errors during the exporting process.

\subsubsection{Verification Results of the Post-Analyzer}
The daily number of hijacks that were not verified by the post-analyzer is illustrated in Figure \ref{fig:invalid_bc}, with a total of unique 5,163 entities \footnote{An entity refers to a unique pair of (origin AS, announced prefix).} placed in quarantine throughout the measurement period.
As indicated by the ``Unverified'' line in the figure, most anomalous routes identified during the four distinct surge periods were verified, supporting the efficacy of this post-verification approach.

To examine the characteristics of hijacking events, we define an event as a series of hijacks originating from a single AS within a day.
6,795 hijacking events were ascertained by the post-analyzer, including 3,163 events initiated by AS212483.
Additionally, 513 ASes exhibited frequent abnormal behavior, resulting in 3,412 events.
For instance, AS395808 was responsible for 73 events during the measurement period. Figure \ref{fig:hijacks_occur_fre_cdf} in Appendix \ref{app:CDFs_events} illustrates the distributions of event occurrences and frequencies for these 513 ASes. We found that 55\% of these ASes initiated events with a frequency of no more than 14 days. We speculate that such frequent anomalous events are likely caused by technical misconfigurations or other inadvertent behavior rather than deliberate malicious activities.


\subsubsection{Whitelist}
\begin{figure}[t!]
\centerline{\includegraphics[width=0.5\linewidth]{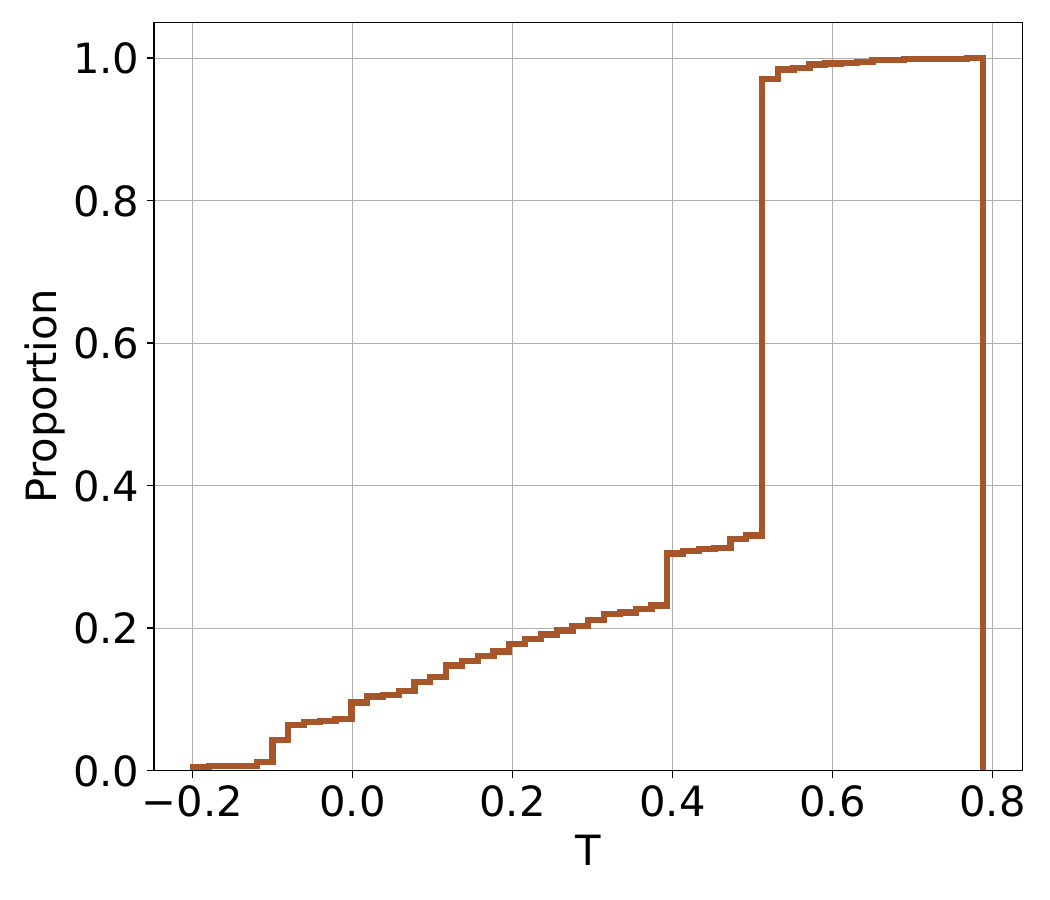}}
\vspace{-15pt}
\caption[CDF of tightness values for benign conflicts.]{\small{CDF of tightness values for benign conflicts identified by the ML-classifier.}}
\label{fig:bc_tightness}
\end{figure}

The benign conflicts identified by the ML-classifier are placed in quarantine for further examination.
We calculate the tightness level between two conflicting origins for each benign conflict using the formula defined in Section \ref{subsec:quarantine}.
Figure \ref{fig:bc_tightness} demonstrates the CDF of tightness values for these benign conflicts.
We specify the $T_{thr}$ value to be 0.3, to ensure both a sufficient quantity of whitelisted items and high reliability of the whitelist.
As a result, 80\% of the identified benign conflicts were whitelisted.
The remaining routes were combined with the unverified routes by the post-analyzer for further behavior monitoring, resulting in 2,321 additional routes added to the whitelist.
In total, the whitelist contains 52,846 items.
\vspace{-10pt}
\subsection{Root Causes of Benign Conflicts} \label{subsec:rootcauses_BC}
We explore the root causes of 52,846 benign conflicts we whitelisted previously. We do not assume that these conflicts are attributed to human errors but explore other four potential factors that may have led to them.
We show the distribution of these factors in Table \ref{tab:benign_conflict_types} and explain them below. Note that a benign conflict can be caused by more than one factor.

\noindent\textbf{Prefix deaggregation.} \label{subsubsec:prefix_deagg}
For the benefit of traffic engineering, ASes have continuing incentives to announce more specific prefixes, which are generated by prefix deaggregation \cite{winter2012explicitly, kalogiros2009understanding}. However, more specific prefixes would result in ROA conflicts due to maximum length mismatches (see \textit{OriginMatch}).
We discovered that 37,904 routes with benign conflicts had an inconsistent length compared to their respective ROAs. This inconsistency may have resulted from prefix deaggregation by ASes.

\noindent\textbf{Dependencies.} Persistent benign conflicts can arise due to downward dependencies on AS paths, where upstream ASes often announce prefixes on behalf of downstream ASes, or vice versa, due to misconfiguration or traffic engineering.
A total of 10,683 whitelisted routes may involve such dependencies. We identified 7,584 benign conflicts with provider-customer relationships, 2,073 with a \textit{Parent} relationship, and 631 with only a relatively greater AS dependency (\textit{Depen}). Additionally, we found 395 routes with an \textit{ASdist} value of 0, but no other relationships between their two distinct origins. These routes were included in this category, as they may implicitly indicate a \textit{Parent} relationship.

We explain the dependencies with one example discovered during our measurements. An announcement for prefix 103.158.51.0/24 originated from AS134697, while the authorized owner of the prefix was AS141403. We contacted the operators of AS141403 and learned that AS134697, the upstream provider of AS141403, had misconfigured their BGP routers during an upgrade process. As a result, instead of propagating BGP routes learned from AS141403, AS134697 advertised routes on behalf of AS141403.
Since AS141403 has issued an ROA for the prefix, networks deploying ROV would discard the routes conflicting with the ROA.
However, AS141403 was unable to resolve the issue with AS134697, as AS134697 did not correct the error but instead promised to reroute the traffic to AS141403. This resulted in a persistent benign conflict.


\noindent\textbf{Multi-origins.} Another common type of benign conflict is a MOAS conflict, as described in Section \ref{sec:lov_features}.
 We found 3,732 benign conflicts where two distinct origins belonged to the same organization.
 
\noindent\textbf{Delayed ROAs.}
This type of benign conflict occurs when operators do not promptly update the RPKI records, especially during the transfer of prefix ownership from one provider to another.
For instance, when a prefix is switched to a new owner, but its ROAs are not updated immediately, this situation may lead to such conflicts.
We identified 3,154 instances where two conflicting sources have no relationships (such as PC or MOAS, etc.), yet verification with the IRRs indicated these routes were valid.
These conflicts reveal a possible delay in updating corresponding ROAs after changes to IRR data.
Upon further analysis, we found that 1,830 out of these conflicts eventually resulted in RPKI valid status when assessed using the most recent ROAs after the measurement period. This provides support for the hypothesis of delayed ROA updates.
\begin{table}[t!]
 \renewcommand{\arraystretch}{1.0}
  \centering
\fontsize{7}{7}
  \begin {tabular}{ccccc}
  \toprule
    \textbf{Deaggregation} & \textbf{Dependencies} & \textbf{Multi-origins} & \textbf{Delayed ROAs} \\
    \midrule
   37,904 & 10,683 & 3,732 & 3,154\\\bottomrule
   \end{tabular}
\caption{\small{Categorization of benign conflicts factors.}}
\label{tab:benign_conflict_types}
\vspace{-15pt}
\end{table}

\vspace{-10pt}
\subsection{Confirmation of Events via Email Surveys} \label{sec:root_causes_events}
Subsequently, we verified the hijacking events through email surveys.
We sent hijack reports to the relevant ASes via email.
Excluding the events attributed to AS212483 and the frequent anomalous events, we were left with 220 hijack events. We sent 194 emails to potential perpetrator ASes and another 194 emails to potential victim ASes. We described the detected events and provided a malicious BGP route as an example in our emails.
We received 10 responses that confirmed the events.
From these responses, we identified potential root causes of hijacking, including technical errors, ownership changes, and non-origin perpetrators (i.e., perpetrators are in the middle of an AS path).
A detailed analysis of these hijacking causes is provided in Appendix \ref{app:root_cause_hijack}.
As mentioned earlier, we manually ensured that confirmed hijacks were not included in our whitelist.

As noted in Section \ref{sec:lov_overview}, the email survey aims to provide evidence for potential human intervention, not to confirm the accuracy of identified hijacks, but to support the creation of a trusted whitelist.
\vspace{-5pt}
\subsubsection{Critical Lessons from Hijacking Events}
A notable point is that all the victims had already issued ROAs. Assuming attackers had no specific targets and accessed the ROAs before engaging in malicious activities, they would intuitively be more inclined to attack ASes that lack ROAs, thereby circumventing ROV's filtering. Despite this, technical misconfigurations, changes in ownership, or other unintended issues can still result in hijacks targeting victims with ROAs.
In other words, hijacks can occur even when ROAs are in place. This challenges the belief among some network operators that issuing ROAs alone is sufficient to protect their networks from hijacks. Since ROV primarily benefits other networks, some operators may choose to issue ROAs without implementing ROV. This reluctance to deploy ROV can hinder its broader adoption across the Internet, which in turn reduces the security benefits of having ROAs. Our findings underscore the importance of not only issuing ROAs but also actively deploying ROV.

\vspace{-5pt}
\section{Key Questions} \label{sec:questions}
This section addresses key questions about \lov.

\textit{Does ML-Classifier need to retrain?}
In this work, we do not retrain the ML classifier with new data, primarily because the features used can help reduce issues with model generalization to newly emerging data, as described in Section \ref{subsubsec:model_general}.
Nevertheless, expanding the size and diversity of the training data could potentially enhance the classifier's effectiveness and adaptability to various scenarios.

\textit{Why is post-analyzer necessary?}
The post-analyzer not only identifies potential benign conflicts mistakenly flagged as hijacks by the ML classifier but also detects possible hijacking events.
The classifier identifies each hijack as an individual instance within an event, without automatically linking them to hijacking events. To associate hijacks with a specific event, the first step is to identify the perpetrator AS that initiated the event, along with the occurrence time. The post-analyzer helps in identifying these initiators and the event time, assisting security analysts in subsequent event analysis.

\textit{Can \lov\ be deployed in individual networks?}
\lov\ can be deployed within a network that has ROV in practice, to create a local whitelist, ensuring that legitimate traffic with benign ROA conflicts is not mistakenly filtered by ROV. This approach removes the reliance on, and potential distrust of, third-party whitelists. However, to prevent any impact on BGP convergence, \lov\ must operate separately from ROV. For instance, \lov\ can function in the background without interfering with the ongoing ROV processes.

Additionally, \lov\ employs various mechanisms, including ML-based, signature-based, heuristic-based, and analyst-driven approaches, which may require security personnel with expertise in areas such as ML technologies. In addition, deploying \lov\ independently may also incur significant costs associated with hardware, software, and maintenance.

We do not recommend directly incorporating the ML classifier into the existing ROV mechanism (as SROV is expected to do). Although this approach might seem straightforward to implement and deploy, the ML-based system is vulnerable to adversarial attacks, which could impair the ROV's ability to detect hijacks.

\textit{What are the differences between SROV and \lov?}
Both SROV and \lov\ incorporate classifiers to distinguish between benign conflicts and real hijacks. The primary difference between the two classifiers is not the use of different technologies—non-ML versus ML—but their design goals. \lov\ aims to preserve as many benign conflicts as possible while maintaining the ROV system's existing ability to prevent hijacks. In contrast, SROV seems to overlook the importance of this balance, leading to an impractical approach. Furthermore, we do not integrate SROV's route duration-based method into \lov\ because it relies on longitudinal route observation, which could affect the whitelist's timeliness. However, during quarantine, \lov\ monitors route behavior (e.g., its duration), complementing the ML classifier.

\textit{What are the differences between \lov\ and BGPmon?}
BGPmon is often considered a reliable monitoring service for BGP anomalies.
\lov\ and BGPmon have a different focus: \lov\ identifies and analyzes benign conflicts, while BGPmon targets the detection of BGP route anomalies such as hijacks. Looking towards the future, there are opportunities for collaboration between \lov\ and BGPmon. For instance, hijacks or events detected by BGPmon could enhance \lov\, such as updating \lov's classifier with new hijacks. Conversely, BGPmon could leverage benign conflicts detected by \lov\ to minimize potential false positives.

Unlike BGPmon, the hijacking events identified by \lov\ are not publicly disclosed. \lov\ focuses exclusively on confirming hijacking events through email surveys to provide evidence for possible human intervention. Consequently, the accuracy in identifying these events has a limited impact on the reliability of the whitelist.

\textit{What are the advantages, limitations, and security considerations of \lov?}
The whitelist provided by \lov\ is easy to update, maintain, and manage, with minimal resource overhead. More importantly, the additional step after ROV enforcement—checking whether the RPKI-invalid route is on the list—incurs negligible time cost, thus minimizing the impact on BGP convergence.
However, behavior monitoring and human intervention in quarantine often require days of route observation, potentially delaying the addition of some benign conflicts to the whitelist and affecting its timeliness. Future work will explore more efficient inspection mechanisms. In rare cases where the victim network is the provider of the attacker's network, the ML classifier may fail to detect the hijack. However, proximity between the victim and malicious networks can often facilitate quicker attack detection and mitigation, provided appropriate monitoring systems are in place. Future work will incorporate additional features to address such attacks.
Additionally, all codes, data, and models are kept confidential and are not publicly available. The operational process of \lov\ is also not transparent.
While these measures can help prevent potential vulnerability exposure to adversaries, ensuring data security and service integrity, they may lead to distrust in the whitelist and hinder its broader adoption. One potential solution might be to involve authoritative organizations (e.g., cybersecurity agencies) to regularly monitor and assess \lov, certifying its effectiveness and reliability. The potential misuse of the whitelist represents another limitation in the deployment of \lov. As previously mentioned, users can access the whitelist through the APIs we provide. Similarly, adversaries might acquire the whitelist, alter it, and distribute a falsified version under the guise of \lov.
One way to mitigate this risk could be to restrict access to the whitelist of benign conflicts to users who verify their identities. For example, users might be required to submit an email request from an organizational address and provide certification of their identity before gaining access.

\vspace{-10pt}
\section{Future Directions} \label{sec:lov_future}

\noindent\textbf{Open questions and improvements.} Our analysis of benign conflicts in Section \ref{subsec:rootcauses_BC}, revealed that approximately 72\% of these conflicts (see "prefix deaggregation") involved prefixes with a matching origin to the ROAs but violated the allowed prefix length.
Since ROV verifies both the origin and the prefix length against the ROAs, these benign conflicts are incorrectly filtered out.
Given that actual hijacks are rare compared to benign conflicts, the current ROV mechanism, with its strict filtering rules aimed at preventing hijacks, may be impractical due to its impact on legitimate routes.
Our observations raise a future research question: Could the filtering strategies in ROV be adjusted to focus exclusively on filtering routes with origin violations against ROAs? This question arises from the fact that most hijacks, such as prefix or sub-prefix hijacks, involve an invalid origin compared to ROAs. While validating prefix length can help filter some hijacks related to AS path manipulation (e.g., when a correct origin AS is appended at the end of the path but a more specific prefix than allowed by the ROA is announced), there are existing path validation mechanisms designed to address these issues, such as BGPsec [RFC8205] and Autonomous System Provider Authorization (ASPA) \cite{aspa}.
Future research will investigate real-world ROV filtering strategies used by ASes, specifically exploring whether they filter all RPKI-invalid routes or allow routes with a matching RPKI origin to pass through.

Benign conflicts with an RPKI-valid origin AS are straightforward to handle and do not necessarily depend on other relationships for identification. This observation inspires us to enhance \lov\ in future work. We will exclude the identification of such benign conflicts from the ML classifier and explore new features to better detect other types of benign conflicts with unknown causes; for example, we will resolve the domain names associated with two conflicting origins and examine the similarity between these domain names.

\noindent\textbf{Future applications.}
Cloud services like AWS, Google Cloud, and Azure offer BYOIP (Bring your own IP), allowing customers to use their own IP addresses for cloud resources. To provision BYOIP, such as in Google Cloud, customers need to create a \textit{public advertised prefix} and an ROA for this prefix that points to the ASN of the cloud service provider.
Google Cloud recommends that customers submit another ROA request for the same prefix and prefix length but with their own ASNs as the origin to avoid benign ROA conflicts when customers need to advertise the prefix \cite{googlecloud}. However, customer behavior is often dynamic—they may ignore this recommendation or fail to maintain the ROA, even if it was initially created. Additionally, when customers deprovision BYOIP, they must wait 14 days before deleting the ROA that authorizes the cloud service provider \cite{googlecloud}.
This delay can also result in benign conflicts as customers might advertise the prefix during the waiting period.
As more users adopt cloud services, BYOIP-related benign conflicts may increase, potentially affecting the user experience. Looking forward, \lov\ has the potential to mitigate this impact. We discuss additional future applications of \lov\ in Appendix \ref{app:add_future_use}.


\section{Conclusions} \label{sec:conclusion}
RPKI plays a crucial role in preventing BGP hijacking. It leverages ROV to ensure that only routes from authorized owners are accepted by routers in BGP announcements.
Despite the effectiveness of RPKI, its deployment has been progressing at a slow pace.
BGP routes with benign origin conflicts with ROAs, so-called benign conflicts, which appear as hijacks but indeed are legitimate, will be identified by ROV as invalid and filtered, leading to the loss of legitimate traffic. This raises concerns among network operators about potential revenue loss and service disruptions, diminishing their motivation to deploy ROV.

In this work, we developed Learning Origin Validation (\lov) to whitelist benign conflicts on an Internet-wide scale, preventing BGP announcements with benign ROA violations from being lost. Network operators can download and install these whitelisted benign conflicts on their ROV-enforcing routers to preserve the affected BGP routes, addressing their concerns. This approach thus has the potential to promote the broader deployment of RPKI and ROV across the Internet.
A longitudinal live measurement demonstrated \lov’s real-world effectiveness, resulting in the whitelisting of 52,846 benign conflicts. Additionally, our investigation identified four factors contributing to benign ROA conflicts beyond human errors. Our observations raise an open question about the current ROV mechanism, potentially motivating further research. Moreover, our discussions indicate that \lov\ holds promise for offering benefits to critical services such as cloud services.

\ignore{
\noindent\textbf{High validity.} LOV first use an ML classification model to effectively distinguish between benign and suspicious announcements, and then use a post-analyzer to further verify whether a suspected malicious announcement is from a real attacker AS, guaranteeing a high accuracy for identifying BGP anomalies. \\

\noindent \textbf{Limitations}
To be continued ...
Discuss about the limitations about post-analyzer (false positive), this method can effectively detect anomalies as anomalous routing behavior emerges, however, in some cases, there will be false alarms, for example, when the ISP has routing changes due to traffic engineering or load-balancing, new routing policies, etc. Or the router feeding the route-views collector restart. 
The location of route-view collectors may affect the global visibility in the routing table, thus degrading the accuracy of post-analyzer.
False positive may be due to the routing instability not because of malicious behavior. If we identify periodical burstiness characteristic of BGP announcements of an AS, we identify it as routing flapping (instability) not malicious behavior. 
A reason
Why route leaks may be detected incorrectly, is that the relationships inferred by CAIDA are derived with an understanding of complex/hybrid relationships, but still output only peer-to-peer and client-to-peer relations \cite{wijchers2014bgp}. 
Hijacking confirmation and diagnosis: we directly contact the corresponding network operators if a stable prefix hijacking is detected. We provide detailed information collected to help their diagnosis work. 
}

\begin{acks}
This work has been co-funded by the German Federal Ministry of Education and Research and the Hessian State Ministry for Higher Education, Research and Arts within their joint support of the National Research Center for Applied Cybersecurity ATHENE and by the Deutsche Forschungsgemeinschaft (DFG, German Research Foundation) SFB~1119.
\end{acks}

\balance


\appendix
\section{CDFs of Occurrences and Frequencies of Benign Conflicts} \label{app:cdf_benign_conflicts}
\begin{figure}[t!]
  \centering
  \begin{minipage}[b]{0.4\linewidth}
    \centering
    \includegraphics[width=\linewidth]{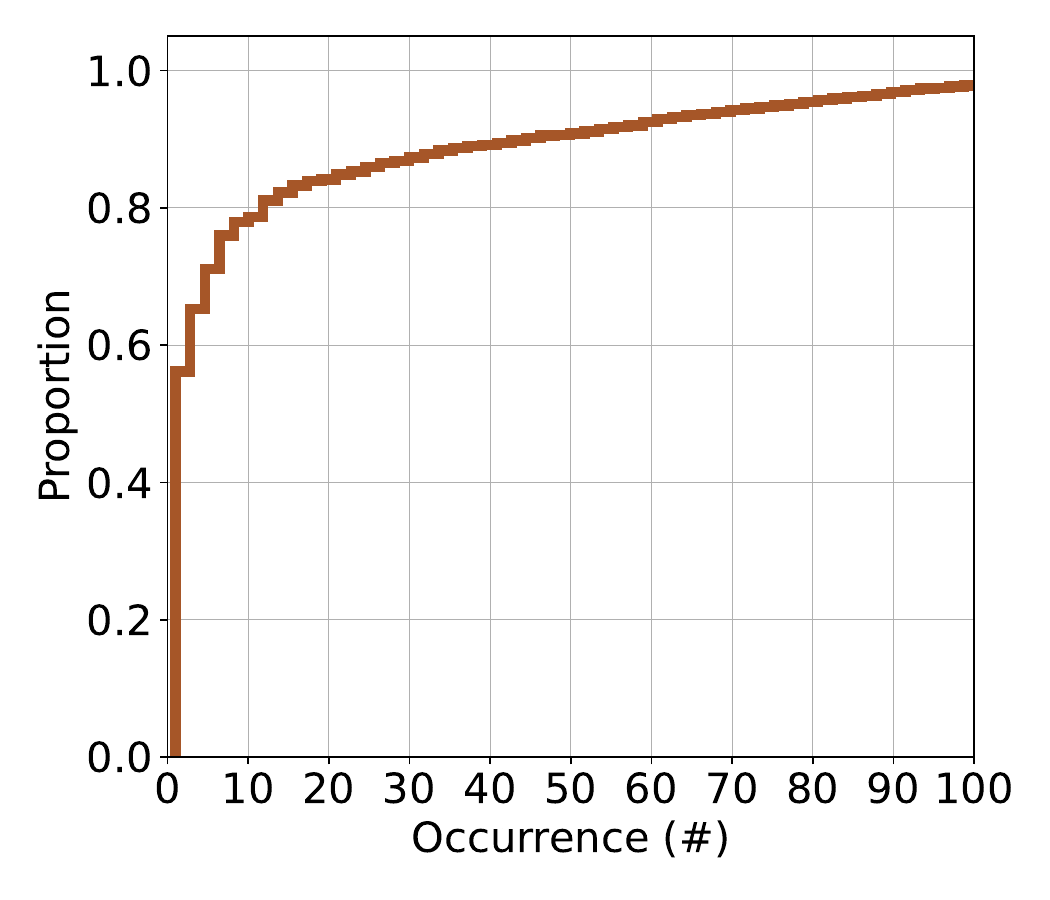}
  \end{minipage}
  \hfill
  \begin{minipage}[b]{0.4\linewidth}
    \centering
    \includegraphics[width=\linewidth]{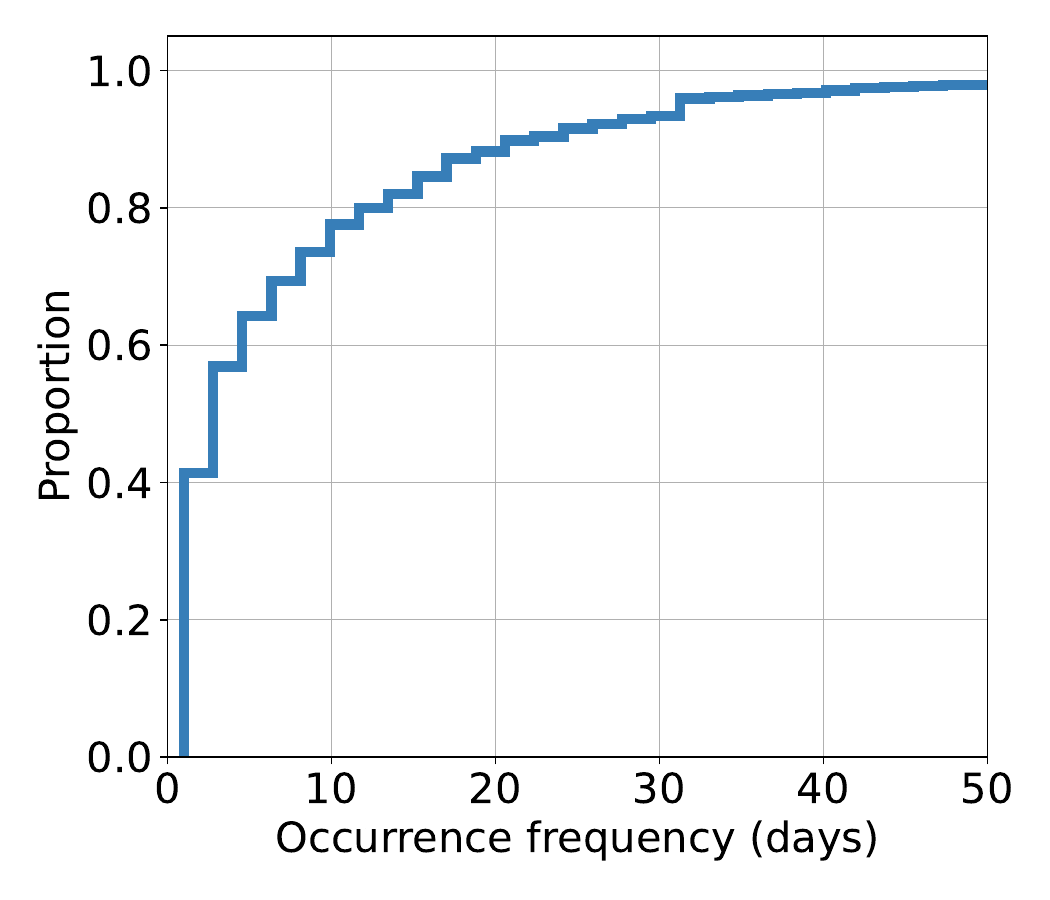}
  \end{minipage}
  \vspace{-10pt}
  \caption{\small CDFs of occurrences and frequencies of benign conflicts.}
  \label{fig:occur_fre_cdf}
\end{figure}

The occurrences and frequencies of these benign conflicts are displayed in Figure \ref{fig:occur_fre_cdf}. An ``occurrence'' refers to the appearance of a particular benign conflict during a day. 
 We observed 33,191 (52\%) benign conflicts that appeared on at least two days during the measurement period, and 80\% of them had an average occurrence frequency of less than 14 days.
 This observation is in line with our expectation that benign conflict tends to be persistent and used actively.
 
\section{Feature Scaling} \label{app:feature_scale}
Compared with other features that fall into [0, 1], $d$ exhibits a larger upper bound. The larger values may cause bias in the classification model. Hence, we scale this feature before training.
In our experiments, we discovered that scaling $d$ into the range of [0, 1] using a commonly used method like minimum-maximum scaling may result in a number of false negatives, i.e., misidentifying hijacks as benign conflicts, negatively impact the quality of the whitelist.
These mis-recognitions often occur when hijacks entail two conflicting origins that are in close proximity.

\begin{figure}[b!]
  \centering

   \begin{minipage}[b]{0.45\linewidth}
    \centering
    \includegraphics[width=\linewidth]{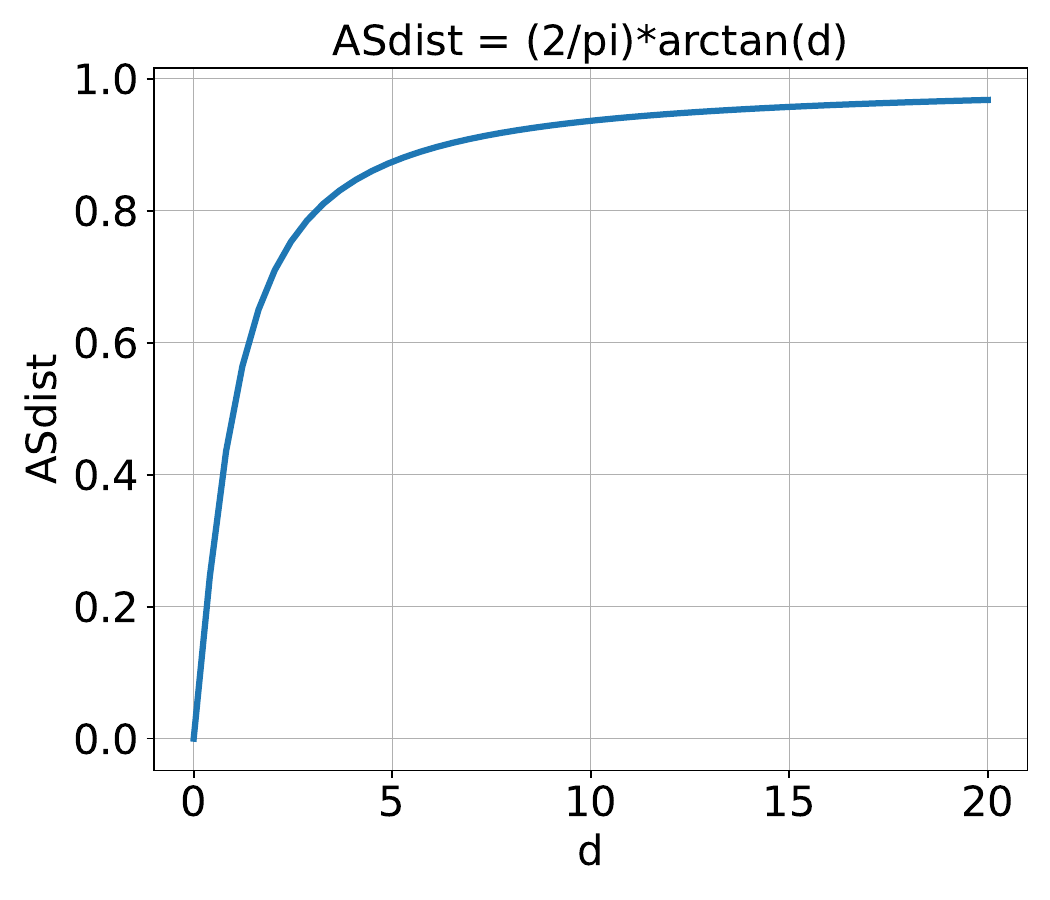}
    \vspace{-25pt}
    \caption{\small Transformation.}
    \label{fig:arctan}
  \end{minipage}
  \hfill
  \begin{minipage}[b]{0.45\linewidth}
    \centering
    \includegraphics[width=\linewidth]{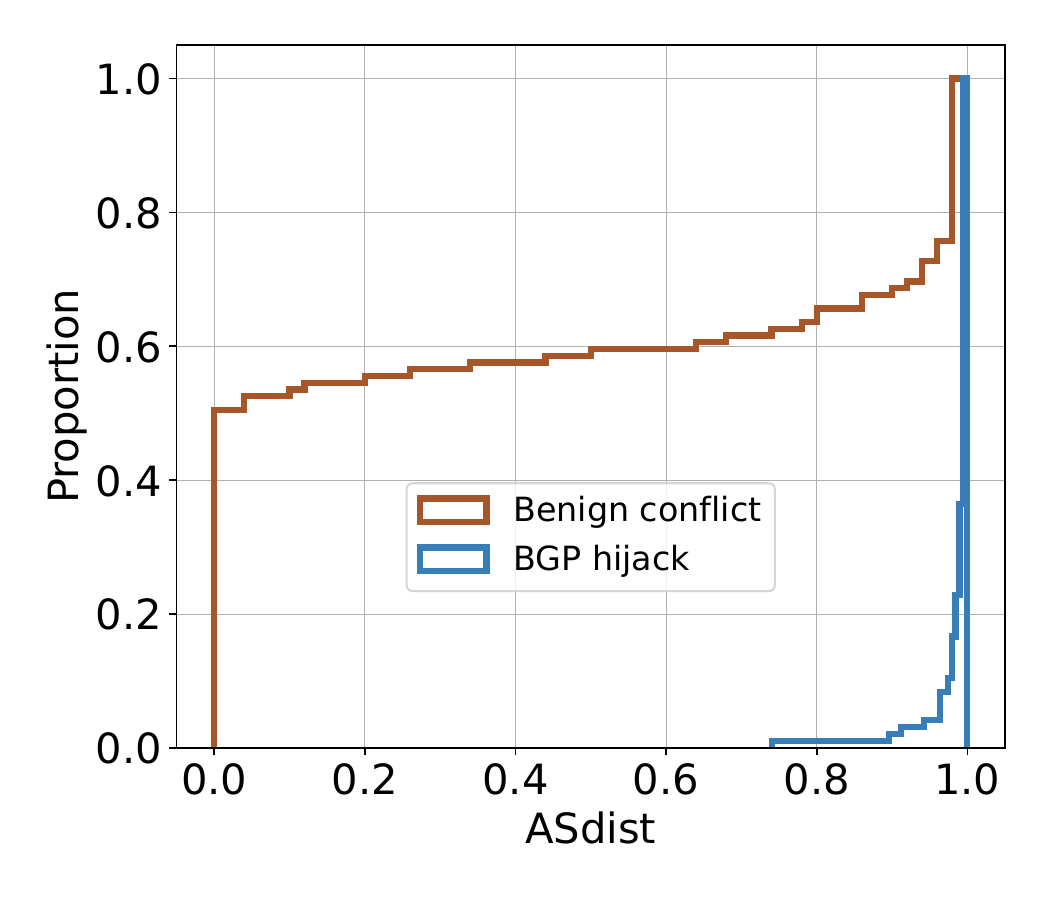}
    \vspace{-25pt}
    \caption{\small CDFs of $ASdist$.}
    \label{fig:ASdist_cdf}
  \end{minipage}
  
  \label{fig:both}
\end{figure}

To address this, we apply the arctangent function to the Euclidean distance ($d$) between the conflicting origin ASes, resulting in a new distance metric $ASdist$ called AS distance, as given by the formula: {$ASdist = \frac{2}{\Pi}arctan(d)$}.
Figure \ref{fig:arctan} shows the transformation curve from $d$ to $ASdist$.
This transformation is specifically chosen for its high sensitivity to hijacks.
Typically, the $d$ value of benign conflicts approaches 0, as presented in Figure \ref{fig:d_cdf}, and we expect larger distance values for hijacks.
As illustrated in Figure \ref{fig:arctan}, small increases in the distance can cause a significant increase in the $ASdist$ value; for instance, increasing d slightly from 0 to 5 results in a substantial rise 90\% in $ASdist$ to 0.9.
This means the $ASdist$ value is highly responsive to increases in distance, making it more effective to detect hijacks (especially for hijacks in close proximity). 
As depicted in Figure \ref{fig:ASdist_cdf}, a large $ASdist$ value can effectively distinguish between hijacks and benign conflicts. 

\section{Grid Search Settings} 
\label{app:grid_search_settings}
The grid search settings for the other four models (DT, SVM, KNN, and RF) are presented in Table \ref{tab:identi_parameters}.

  \begin{table}[t!]
    \centering
    \fontsize{6}{6}\selectfont
    \renewcommand{\arraystretch}{1.2}
    
    \begin{tabular}{cccc}
        \toprule
        \textbf{Model} & \textbf{Parameter} & \textbf{Search range} & \textbf{Step size} \\
        \midrule
        \multirow{2}{*}{DT} & $max\_depth$ ($d$) & [3, 20] & 2 \\
        & $min\_samples\_leaf$ ($l$) & [2, 20]  & 2 \\
        SVM & $C$ & [0.01, 0.1, 1, 10, 100, 1000, 10000] & - \\
        KNN & $n\_neighbors$ ($k$) & [1, 20] & 2\\
        \multirow{2}{*}{RF} & $max\_depth$ ($d$) & [3, 20] & 2 \\
        & $min\_samples\_leaf$ ($l$) & [2, 20]  & 2 \\
        \bottomrule
    \end{tabular}
    \caption{\small Grid search settings for the primary parameters of the DT, SVM, KNN, and RF models, including the search ranges and step sizes. $max\_depth$ means the maximum depth of decision trees. $min\_samples\_leaf$ represents the minimum number of samples in a leaf node. $C$ refers to a penalty parameter to control error tolerance. $n\_neighbors$ refers to the number of the closest neighbors used for predictions.}
    \label{tab:identi_parameters}
\end{table}

\section{Statistics of Hijacks During Four Surge Period} \label{app:statistics}
 We present the statistics of hijacks identified during four surge periods in Table \ref{tab:specific_days_res}.

\begin{table*}[t!]
   \renewcommand{\arraystretch}{1.2}
   \fontsize{8}{8}\selectfont 
  \centering
  \begin {tabular}{cccccccc}
  \toprule
    \textbf{Dates} & \textbf{Hijacks (\#)} & \textbf{Attackers (\#)} & \textbf{Prefixes (\#)} & \textbf{Victims (\#)} & \textbf{Att$_{m}$} & \textbf{Vic$_{m}$} & \textbf{Peer$_{m}$} \\
    \midrule
    
   2022-10-26 to 2022-10-29 & 14,636 & 3,827 & 5,438 & 328 & AS396982 (191) & AS11351 (2,008) & AS212483 (14,026)    \\
    2022-11-07 to 2022-11-08 & 7,086 & 450 & 5,990 & 243 & AS396982 (4,017) & AS8070 (3,402) & AS212483 (6,719)  \\
    2022-11-16 to 2022-11-21 & 7,940 & 3,185 & 2,363 & 389 & AS36351 (102) & AS8075 (1,183) & AS212483 (7,298) \\
    2022-11-23 to 2022-11-30 & 84,577 & 26,659 & 22,725 & 699 & AS13335 (491) & AS6128 (16,790) & AS212483 (83,916)  \\
   \bottomrule
   \end{tabular}
   \caption[Hijacks identified during four surge periods.]{\small Hijacks identified during four periods. "Hijacks", "Attackers", "Prefixes" and "Victims" refer to the number of detected hijacks, involved perpetrator ASNs, announced prefixes, and affected ASNs, respectively. "Att$_{m}$", "Vic$_{m}$" and "Peer$_{m}$" indicate the attacker ASN who initiated the most hijacks, the victim ASN who was affected by the most hijacks, and the peer ASN who fed the most hijacks to the RouteViews collector (corresponding hijacks are enclosed in brackets). 
   }
\label{tab:specific_days_res}
\end{table*}

\section{CDFs of Occurrences and Frequencies for Events}\label{app:CDFs_events} 
\begin{figure}[b!]
  \centering
  \begin{minipage}[b]{0.43\linewidth}
    \centering
    \includegraphics[width=\linewidth]{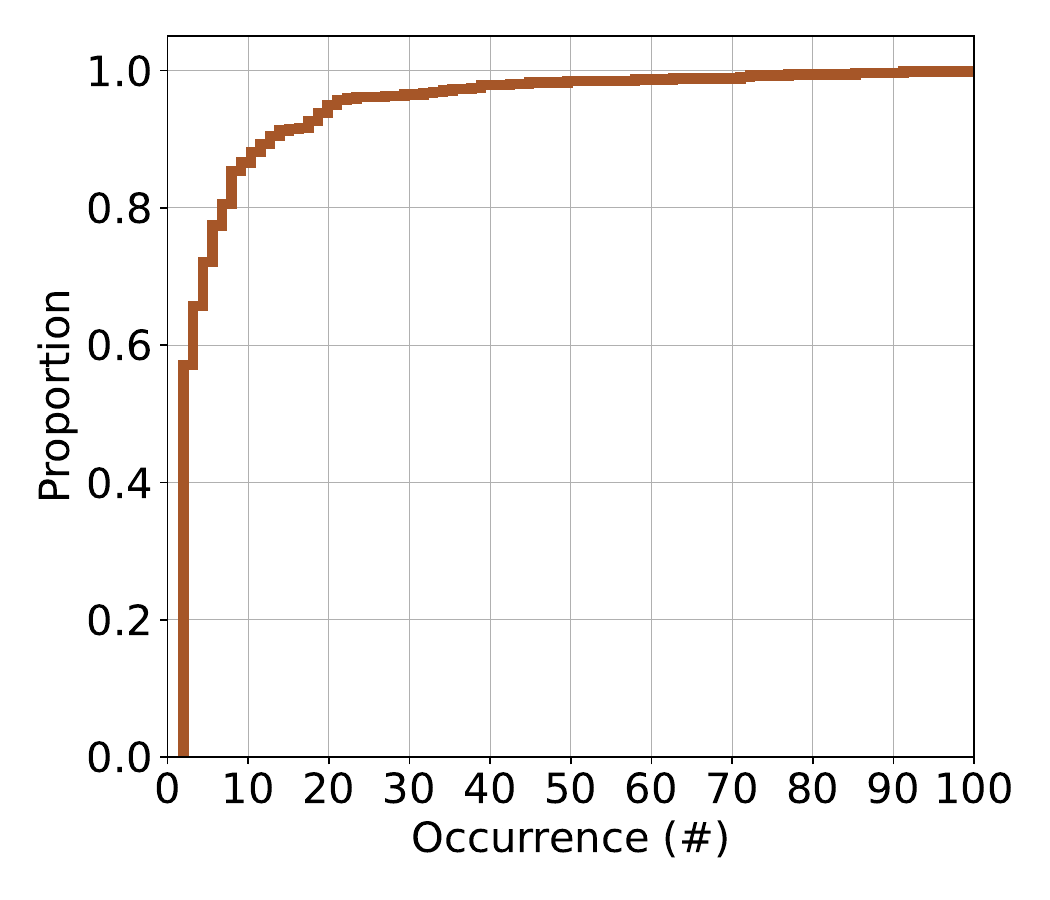}
  \end{minipage}
  \hfill
  \begin{minipage}[b]{0.43\linewidth}
    \centering
    \includegraphics[width=\linewidth]{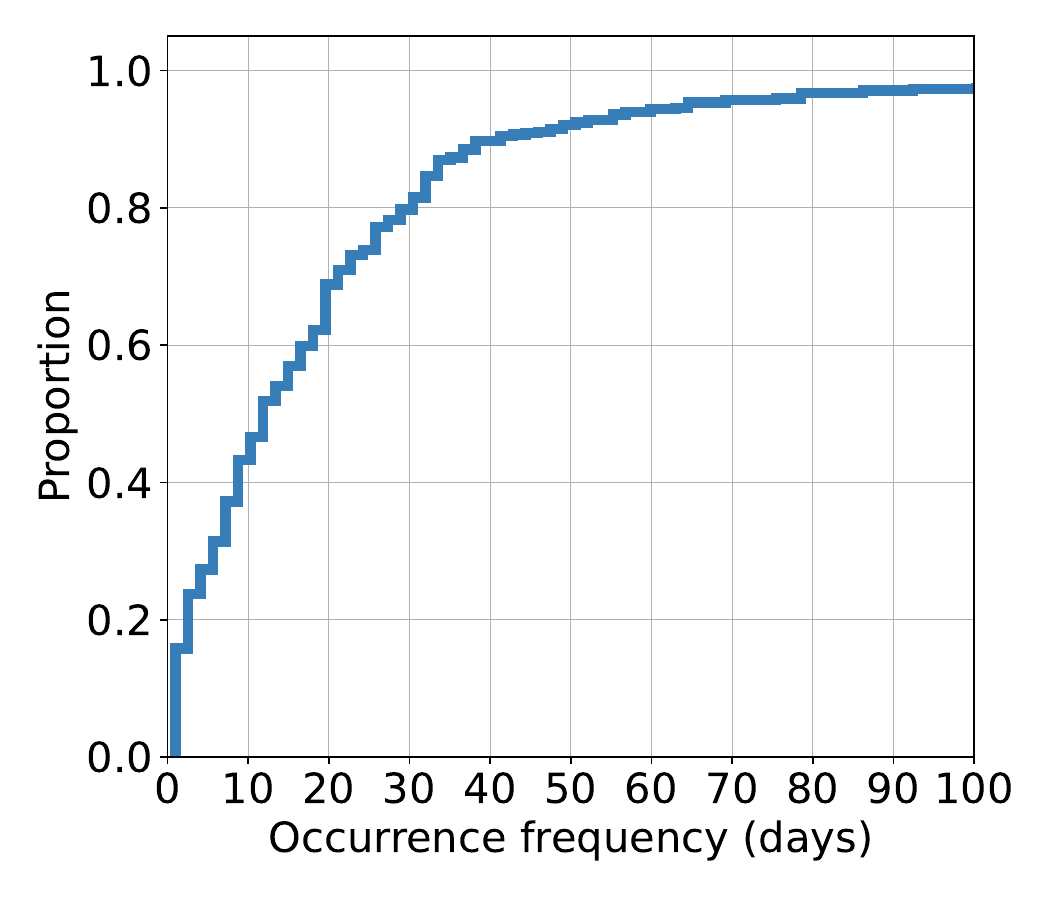}
  \end{minipage}
  \vspace{-10pt}
  \caption{\small CDFs of event occurrences and frequencies for 513 ASes.}
  \label{fig:hijacks_occur_fre_cdf}
\end{figure}

Figure \ref{fig:hijacks_occur_fre_cdf} illustrates the distributions of event occurrences and frequencies for 513 ASes
that frequently initiated events.

\section{Root Causes of Hijacking Events} \label{app:root_cause_hijack}
We analyze the hijacking events identified through email surveys, with a focus on investigating their underlying causes.
To achieve this, we sent hijack reports to the relevant ASes via email.
Excluding the events attributed to AS212483 and the frequent anomalous events, we were left with 220 hijack events. We sent 194 emails to potential perpetrator ASes and another 194 emails to potential victim ASes and received 10 responses that confirmed the events.
The factors causing these hijacks are described below. \\

 \noindent\textit{Technical errors.} Six events were due to technical issues. For example, AS52195 who owned 46.149.208.0/20, announced an invalid route to the prefix 20.46.149.208/32. The network operators of AS52195 clarified that this anomaly was potentially caused by a technical error as the announced prefix 20.46.149.208/32 can be generated by a circular byte right shifting on 46.149.208.0/20. \\
 
\noindent\textit{Owner changes.}
We identified two events arising from owner changes. On March 31, AS210837 announced the route to the prefix 45.151.234.0/24, despite AS12695 being the authorized origin according to ROAs. Upon contacting AS12695, we learned that they started announcing this prefix on March 30. Given that AS210837 was the previous owner of this prefix, the observed hijack likely occurred due to a delay on their part in updating the configuration of BGP routers.\\

\noindent\textit{The perpetrator AS is in the middle.}
In our previous study of events related to AS212483, we found that BGP anomalies could be initiated not only by the origin AS. In one of the other two examples, AS134196 was identified as hijacking the prefix 24.103.24.0/23, which belonged to AS12271. The BGP path of the hijack was [..., AS3223, AS55933, AS134196]. When we contacted AS134196, they believed the routing path was abnormal because they expected AS3223 to be their immediate upstream provider for network traffic. We then investigated the announcements from AS55933 on the same day and found that AS55933 had also originated routes for this prefix.
This indicated that AS55933 was highly suspected of initiating the event, rather than the origin AS in the path.

\section{Addressing Challenges in ML Application} \label{app:ml_challenges}
We then discuss the key challenges encountered in developing and deploying the ML-based classifier, the solutions implemented, and potential solutions for future work.

\subsection{Ground Truth Data}
In collecting ground truth data for developing the ML-classifier, we leverage the long-duration nature of benign routes to identify and extract benign conflicts from RPKI-invalid BGP routes. While this approach helps ensure that the collected benign conflicts are likely benign, it may miss short-lived conflicts, such as those lasting only a few hours or days. Consequently, the absence of transient benign conflicts in the ground truth data could impact data diversity and, in turn, affect model performance.

Recall that we collected BGP hijacks from BGPmon. Although BGPmon is highly regarded as a reliable source, it still encounters issues with false positives. To ensure the cleanliness of the collected data, we performed pre-processing and filtered out instances mistakenly identified as hijacks. However, due to the lack of transparency in BGPmon’s operational process, we cannot guarantee that the remaining BGP hijacks are genuinely malicious.

A potential improvement would be to incorporate both benign conflicts and BGP hijacks confirmed through email surveys to retrain the ML-classifier in future work.

\subsection{Model Performance and Reliability}
Multiple ML models are used to implement the classifier. Their effectiveness is evaluated and compared to select the optimal model for identification. Cross-validation and grid search are also employed to achieve the best performance.

Additionally, we introduce a post-analysis mechanism to verify the hijacks identified, aiming to correct any errors where the ML-classifier might misidentify benign conflicts as hijacks. Notably, benign conflicts flagged by the classifier are not immediately deemed benign but are placed in quarantine for further review before whitelisting. This approach helps mitigate the impact of the classifier's failure to detect hijacks on the whitelist.

As mentioned in Section \ref{subsubsec:robustness}, the ML-based classifier may be vulnerable to adversarial attacks, where crafted hijacks could evade the classifier's detection and contaminate the whitelist.
Our quarantine mechanism for further inspection can mitigate such a risk, ensuring the reliability of the whitelist.

\subsection{Model Generalization}
As mentioned in the previous section, our feature set focuses solely on capturing the relationships between two conflicting origins rather than the behavior patterns underlying the routes.
This can mitigate the impact of the evolving behavior of anomalous routes over time on identification, thereby minimizing the model generalization issues to newly emerging data.

Additionally, our evaluation of new instances supports the model's ability to generalize across different scenarios. Even if the model incorrectly identifies a route, this issue can be addressed in subsequent analysis, including post-verification of hijacks and further review of quarantined routes.

\subsection{Model Interpretability}
The inherent opaqueness of ML models can obscure their decision-making processes. We also quantify the importance of each feature in the identification.
This analysis assists in comprehending the inner workings of the model, enhancing the interpretability of its outputs.

\section{Additional Future Applications} \label{app:add_future_use}
As previously discussed, prefix hijacking presents serious risks including data interception, traffic redirection to malicious destinations, and blackholing of legitimate traffic. These threats pose significant security concerns for various Internet infrastructure services. Adversaries can exploit prefix hijacking to disrupt critical services such as DNS, Web, email, and NTP. For instance, attackers can reroute DNS, HTTP, or NTP requests to servers under their control, enabling them to deliver fraudulent responses to unsuspecting clients.
RPKI's ROV can offer an effective defense against prefix hijacking, protecting these services from such attacks.

However, ROV also filters BGP announcements with benign conflicts.
Networks hosting open DNS resolvers, web servers, email servers, or NTP servers may enforce ROV, or ROV enforcement may be present along the path to these networks. In such cases, clients might inadvertently originate BGP routes with ROA violations for their IP addresses, which ROV flags as invalid and discarded.
As a result, server responses fail to reach clients due to missing routing information, leading to service unavailability.
Conversely, if networks hosting these critical servers initiate announcements with conflicting origins, ROV filtering can affect the visibility of their IP addresses across the Internet, causing service unreachability for clients.

Similar issues can also arise within Cloud environments, potentially impeding users' access to their critical data, such as banking information, stored in the Cloud.
Furthermore, IoT devices with RPKI-invalid IP addresses may face significant communication problems with other IoT devices and central IoT servers, potentially leading to malfunctions in IoT applications.
Disruptions stemming from ROV filtering benign conflicts could lead to trust erosion, reputational damage, and economic losses for Cloud or IoT service providers.

Looking ahead, the widespread deployment of ROV could exacerbate the impact of benign conflicts on various services within the Internet. Adopting \lov\ to prevent ROV from filtering benign conflicts can help mitigate this impact, ensuring the availability and integrity of these services.
\end{sloppypar}

\end{document}